\documentclass[preprint,12pt]{elsarticle}

\usepackage{graphicx}

\usepackage{soul}
\usepackage{color}

\usepackage{amssymb}
\usepackage{amsthm,amsmath}

 \usepackage{lineno}

\journal{arXiv}

\begin{document}

\begin{frontmatter}



\title{Quantification of propidium iodide delivery with millisecond electric
pulses: A model study}


\author[a]{Miao Yu}
\author[a]{Hao Lin\corref{cor1}}
\address[a]{Mechanical and Aerospace Engineering,
Rutgers, The State University of New Jersey, 98 Brett Road, Piscataway,
NJ 08854, USA}
\ead{hlin@jove.rutgers.edu}
\cortext[cor1]{Corresponding Author. Tel.: +1 848 445 2322; fax: +1 732 445 3124.}

\begin{abstract}
A model study of propidium iodide delivery with millisecond electric
pulses is presented; this work is a companion of the experimental
efforts by Sadik et al. \cite{Sadik13}. Both membrane permeabilization
and delivery are examined with respect to six extra-cellular conductivities.
The transmembrane potential of the permeabilized regions exhibits
a consistent value, which corresponds to a bifurcation point in the
pore-radius-potential relation. Both the pore area density and membrane
conductance increase with an increasing extra-cellular conductivity.
On the other hand, the inverse correlation between propidium iodide
delivery and extra-cellular conductivity as observed in the experiments
is quantitatively captured by the model. This agreement confirms that
this behavior is primarily mediated by electrophoretic transport during
the pulse. The results suggest that electrophoresis is important even
for the delivery of small molecules such as propidium iodide. The
direct comparison between model prediction and experimental data presented
in this work helps validate the former as a robust predictive tool
for the study of electroporation.
\end{abstract}


\end{frontmatter}



\newpage

\section{Introduction}

Electroporation is a widely-used technique to deliver active agents
into biological cells and tissue \cite{Neumann_1982,Mir_2000,Gehl_2003,Andre_2004,Schoenbach_2004,Driessche_2005,Costa_2007,Teissie_2009,Teissie_2012}.
The process includes two basic aspects. In the first, the application
of an electric pulse permeabilizes the membrane to gain access to
the cytoplasm \cite{Chang_1990,Wilhelm_1993,Spassova_1994,Djuzenova_1996,Wegner_2011,
Napotnik_2012,Krama_2012,Barnett_1991,Weaver_1996,Neu_1999,Leontiadou_2004,Tarek_2005,Levine_2010,Jianbo_2010}.
In the second, molecules are transported into the cell via mechanisms
such as electrophoresis, diffusion, and endocytosis \cite{Andre_2004,sukharev_1992,prausnitz_1995,Mir_1998,Rols_1998,Faurie_2004,Satkauskas_2002,
Satkauskas_2005,Pavselj_2005,Liu_2006,Pucihar_2008,Kanduser_2009,Miklavcic_2010,
Pavlin_2010_2,Smith_2012,Vasquez_2012,Zimmermann_1990,Wu_2011,Lin_2011,Rosazza_2012}.
In an earlier work by one of us (HL \cite{Sadik13}, henceforth denoted
as Sadik13), we used time- and space-resolved fluorescence microscopy
to quantify the second aspect, namely, the transport of small molecules
via electroporation. This study aimed to differentiate contributions
to total delivery by the various pertinent mechanisms. In addition,
it also provided quantitative data to help interpret trends observed
in earlier experiments, namely, the inverse correlation between delivery
and extra-cellular conductivity {\cite{Djuzenova_1996,Muller_2001}}.
The current work complements the experimental efforts of Sadik13 with
a model study.

The model couples the asymptotic Smoluchowski equation (ASE) \cite{Neu_1999,Neu_2003,Krassowska2007}
for membrane permeabilization with the Nernst-Planck equations for
ionic transport \cite{Jianbo_2011,Jianbo_2012}. {(A list of abbreviations is given in Table \ref{tab:Abbreviations.}.)}
Following Sadik13, the delivery of propidium iodide (PI) into 3T3
mouse fibroblast cells is simulated. The extra-cellular conductivity
is varied between 100 and 2000 $\mu$S/cm. The simulation provides
detailed, dynamic predictions that were not directly measured by the
experiments, including the systematic behavior of the transmembrane
potential (TMP), the membrane conductance, and the pore area density
(PAD). On the other hand, the results on PI delivery is compared directly
with data from Sadik13. This comparison not only validates the numerical
model, but also helps tackle the basic physical processes involved
in electroporation-mediated molecular delivery. 
\begin{table}
\begin{tabular*}{1\linewidth}{@{\extracolsep{\fill}}ll}
\hline 
Abbreviation & Definition\tabularnewline
\hline 
ASE & asymptotic Smoluchowski equation\tabularnewline
FASS & Field-Amplified Sample Stacking\tabularnewline
PAD & pore area density\tabularnewline
PI & propidium iodide\tabularnewline
Sadik13 & reference \cite{Sadik13}\tabularnewline
TFI & total fluorescence intensity\tabularnewline
TMP & transmembrane potential\tabularnewline
\hline 
\end{tabular*}\caption{Definition of abbreviations.\label{tab:Abbreviations.}}
\end{table}

\section{Model formulation}

\begin{figure}
\center\includegraphics[width=0.6\linewidth]{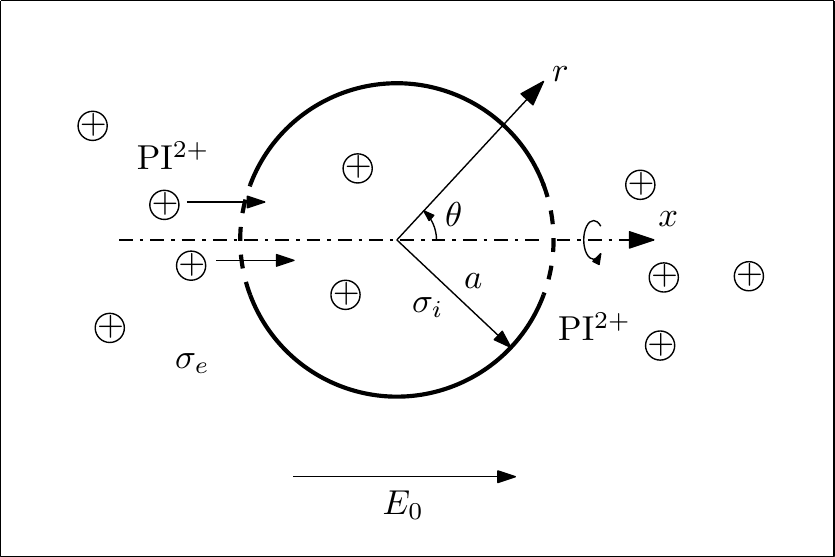}\caption{A schematic of the problem. ($r$, $\theta$) denotes the spherical
coordinate system. $x$ is the axis of rotation, and is aligned with
the direction of field application. The field strength is denoted
by $E_{0}$. The intra- and extra-cellular conductivities are denoted
by $\sigma_{i}$ and $\sigma_{e}$, respectively. \label{fig:schematic}}
\end{figure}

A schematic of the problem is presented in Fig. \ref{fig:schematic}.
A constant pulse with the strength of $E_{0}$ is applied, and axisymmetry
is assumed with respect to the direction of the electric field. A
spherical coordinate system ($r,\theta$) is adopted, and the cell
radius is $a$. The axis of symmetry is denoted by $x$, which is
also the coordinate along the cell centerline. The intra- and extra-cellular
conductivities are denoted by $\sigma_{i}$ and $\sigma_{e}$, respectively.
The PI molecule is a charged ion with a valence number of +2.

The model framework follows that presented in earlier work \cite{Krassowska2007,Jianbo_2011,Jianbo_2012}.
Specifically, the permeabilization model (including the electrical problem) follows Krassowska and Filev \cite{Krassowska2007};
the transport model was developed by the current authors  \cite{Jianbo_2011,Jianbo_2012}. A brief summary is presented below.
For reference, detailed definitions of pertinent variables are given in Appendix A.

Briefly speaking, the Ohmic equations are solved for the intra-cellular
electric potential, $\Phi_{i}$, and the extra-cellular electric potential,
$\Phi_{e}$:
\begin{equation}
\nabla \cdot (\sigma_{i,e} \nabla\Phi_{i,e})=0. \label{eq:Ohmic1}
\end{equation}
 On the membrane, the current density continuity condition is
applied:

\begin{equation}
-\mathbf{n}\cdot\sigma_{i}\nabla\Phi_{i}=-\mathbf{n}\cdot\sigma_{e}\nabla\Phi_{e}=C_{m}\frac{\partial V_{m}}{\partial t}+j_{p},\label{eq:Ohmic}
\end{equation}
where $\mathbf{n}$ is the local unit vector normal to the membrane,
$C_{m}$ is the membrane capacitance, and $j_{p}$ is the local ionic
current density across electropores. The TMP (denoted by $V_{m}$)
is the potential difference across the infinitesimally-thin membrane.
Equations (\ref{eq:Ohmic1}, \ref{eq:Ohmic}) are coupled with the ASE for membrane permeabilization
to track the evolution of both the electric potential and pore statistics.
\begin{equation}
\frac{dN}{dt}=\alpha e^{(V_m/V_{ep})^2}\left( 1-\frac{N}{N_0 e^{q(V_m/V_{ep})^2}} \right), \label{eq:dNdt}
\end{equation}

\begin{equation}
\frac{dr_j}{dt}=U(r_j,V_m,\tau). \label{eq:drjdt}
\end{equation}
Here $N(t,\theta)$ is the local pore number density. $\alpha$, $N_0$, $q$ and $V_{ep}$ are constants. $U$ is the advection velocity, and $\tau$
is an effective membrane tension. According to Krassowska and Filev \cite{Krassowska2007}, pores nucleate at an initial radius, $r_* = 0.51$ nm, and at a rate
described by Eq. (\ref{eq:dNdt}). They then evolve in size according to Eq. (\ref{eq:drjdt}), where $r_j$ is the pore radius. Once Eqs. (\ref{eq:dNdt}, \ref{eq:drjdt})
are solved, the current density through the pores, $j_p$, can be calculated, and used in Eq. (\ref{eq:Ohmic}) (see Eqs. (\ref{eq:jp}, \ref{eq:ip}) in Appendix A). Based on
the pore statistics according to Eqs. (\ref{eq:dNdt}, \ref{eq:drjdt}), we can also compute the PAD, $\rho_p$:
\begin{equation}
\rho_{p}(t,\theta)=A_{p}(t,\theta)/\Delta A,
\end{equation}
where $\Delta A$ is a local area element \cite{Krassowska2007},
and $A_{p}$ is the total area occupied by the pores thereon (see Eq. (\ref{eq:rho_p}) in Appendix A). The
PAD therefore represents a quantification of the degree of membrane
permeabilization. The effective membrane conductance, $g_{m}$, is calculated
by taking the ratio of the local ionic current density and the TMP:
\begin{equation}
g_{m}(t,\theta)=j_{p}/V_{m}.\label{eq:gm-def}
\end{equation}

Three molecular species are considered in this study, which follow the reactive kinetics
\begin{equation}
\mbox{P\ensuremath{\mbox{I}^{2+}}}+\mbox{B}\underset{k_{-}}{\overset{k_{+}}{\rightleftharpoons}}\mbox{PIB}.\label{eq:PIB reaction}
\end{equation}
Here $\mbox{P\ensuremath{\mbox{I}^{2+}}}$ denotes the free PI ion,
B denotes the binding sites in the cytoplasm, and PIB is the compound
which is responsible for the experimentally observed fluorescence
emission. $k_{+}$ and $k_{-}$ are the association and dissociation
constants, respectively. The species concentrations are denoted by
$\mbox{[\mbox{P\ensuremath{\mbox{I}^{2+}}}]}$, $[\mbox{B}]$, and
$[\mbox{PIB]}$, respectively, and obey the Nernst-Planck equations for ionic transport:
\begin{equation}
\frac{\partial X}{\partial t}=\nabla \cdot (\omega FzX\nabla \Phi)+\nabla \cdot (D \nabla X)+\dot{R}. \label{eq:transport}
\end{equation}
Here $X$ is $\mbox{[\mbox{P\ensuremath{\mbox{I}^{2+}}}]}$, $[\mbox{B}]$, or $[\mbox{PIB]}$, $F$ is Faraday constant. 
$\omega$, $z$, and $D$ are the electrophoretic mobility, valence number and molecular diffusivity of the species considered, respectively. 
$\dot{R}$ is a source term due to chemical production according to (\ref{eq:PIB reaction}). Equation (\ref{eq:transport}) is solved for each species
both within and outside the cell, subject to a flux condition on the membrane.
A complete description of the model, including the numerical implementation,
as well as the boundary and initial conditions, are found in \cite{Jianbo_2011,Jianbo_2012},
and are not presented here for brevity. Model parameters specific
to the current problem are listed in Table \ref{tab:List-of-model},
which follows the experimental conditions in Sadik13. Note that in
particular we adopted a value of 0.16 V for $V_{ep}$, the characteristic
voltage of electroporation. The value defers from that used in previous
work \cite{Krassowska2007,Jianbo_2011,Jianbo_2012}, and is determined
from a comparison between experimental data and model simulation in
our recent study \cite{Sadik_2013_2}.

\begin{table}
\begin{tabular*}{1\linewidth}{@{\extracolsep{\fill}}lll}
\hline 
Symbol & Definition & Value/Source\tabularnewline
\hline 
$E_{0}$ & applied field strength & 0.8 kV/cm \cite{Sadik13}\tabularnewline
$t_{p}$ & pulse length & 100 ms \cite{Sadik13}\tabularnewline
$a$ & cell radius & 7 $\mu{\rm m}$ \cite{Sadik13}\tabularnewline
$\sigma_{e}$ & extra-cellular conductivity & 100-2000 $\mu$S/cm \cite{Sadik13}\tabularnewline
$\sigma_{i}$ & intra-cellular conductivity & 4000 $\mu$S/cm \cite{Jianbo_2012}\tabularnewline
$[\mbox{\mbox{P\ensuremath{\mbox{I}^{2+}}}}]_{e,o}$ & initial extra-cellular concentration of $\mbox{P\ensuremath{\mbox{I}^{2+}}}$ & 100 $\mu{\rm M}$ \cite{Sadik13}\tabularnewline
$[\mbox{B}]_{i,o}$ & initial intra-cellular concentration of B & 6.93 ${\rm mM}$ \cite{Jianbo_2012}\tabularnewline
$V_{ep}$ & characteristic electroporation voltage & 0.16 V \cite{Sadik_2013_2}\tabularnewline
\hline 
\end{tabular*}\caption{List of model parameters.\label{tab:List-of-model}}
\end{table}

\section{Results}

In the following, we first present simulated results on the effect
of extra-cellular conductivity on membrane permeabilization. The results
on PI delivery are then presented and compared with experimental data
from Sadik13. For all cases, a single pulse of 0.8 kV/cm and 100 ms
is applied. 

\begin{figure}
\includegraphics[width=0.5\linewidth]{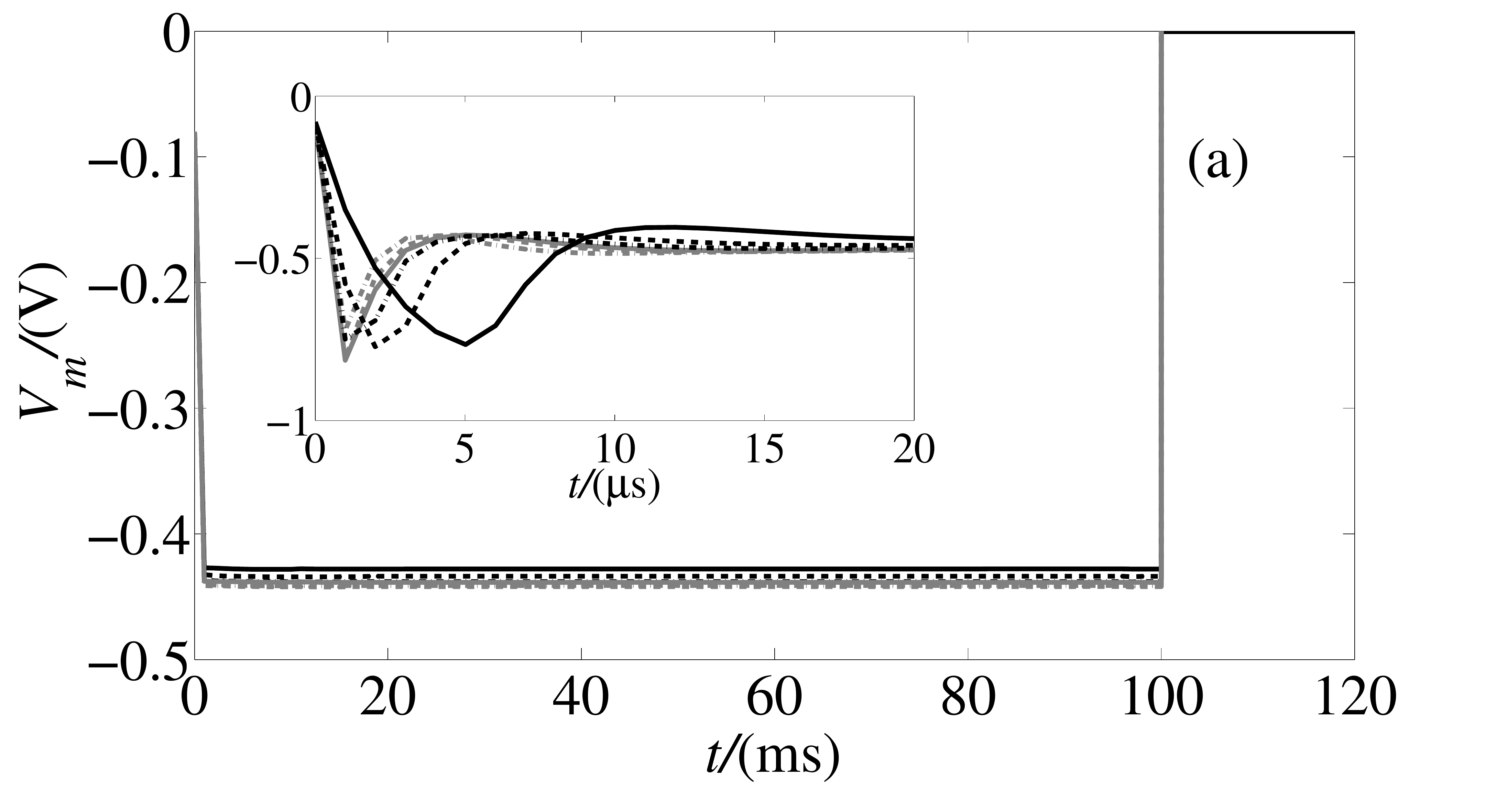}\includegraphics[width=0.5\linewidth]{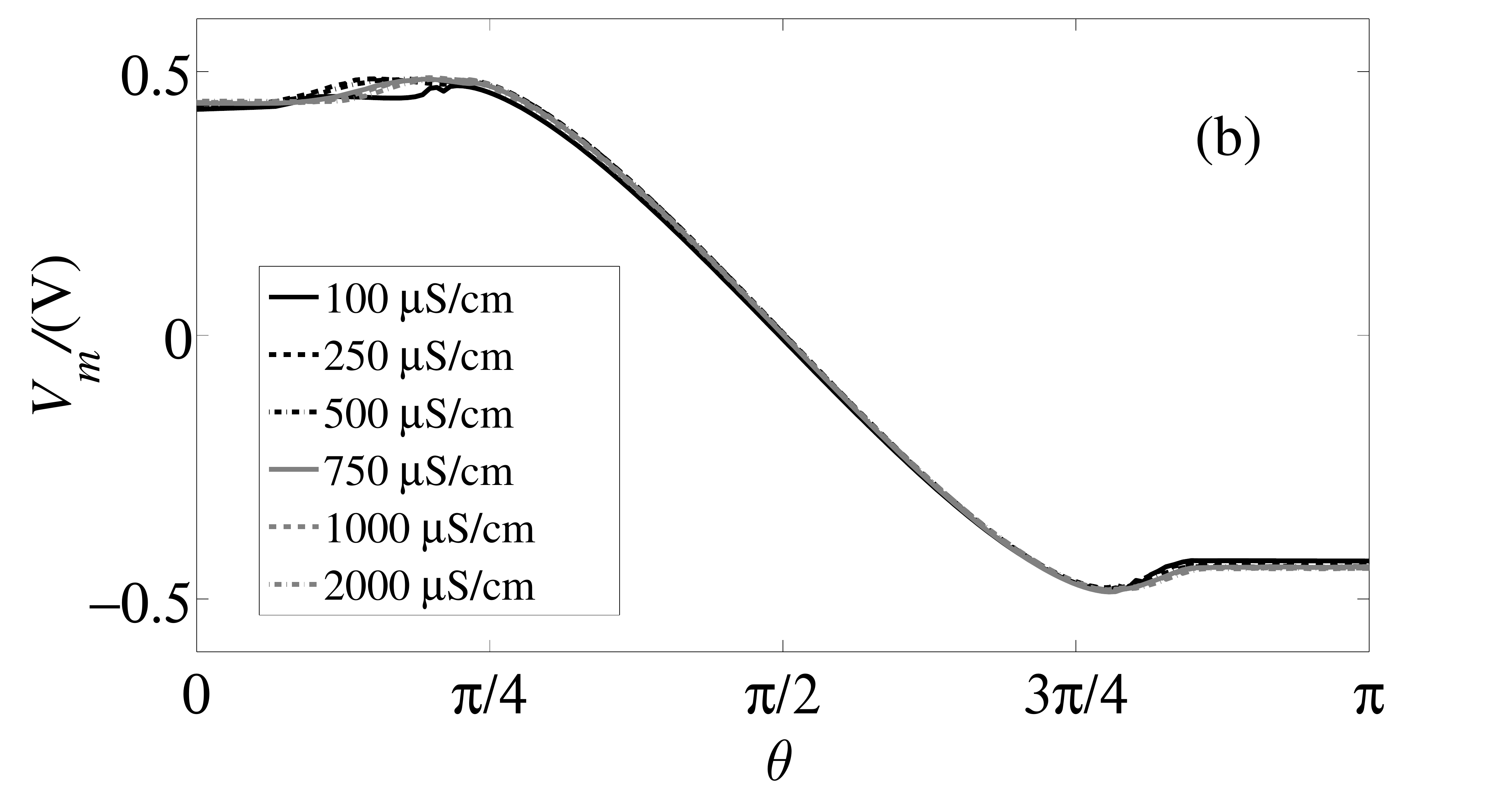}\\\includegraphics[width=0.5\linewidth]{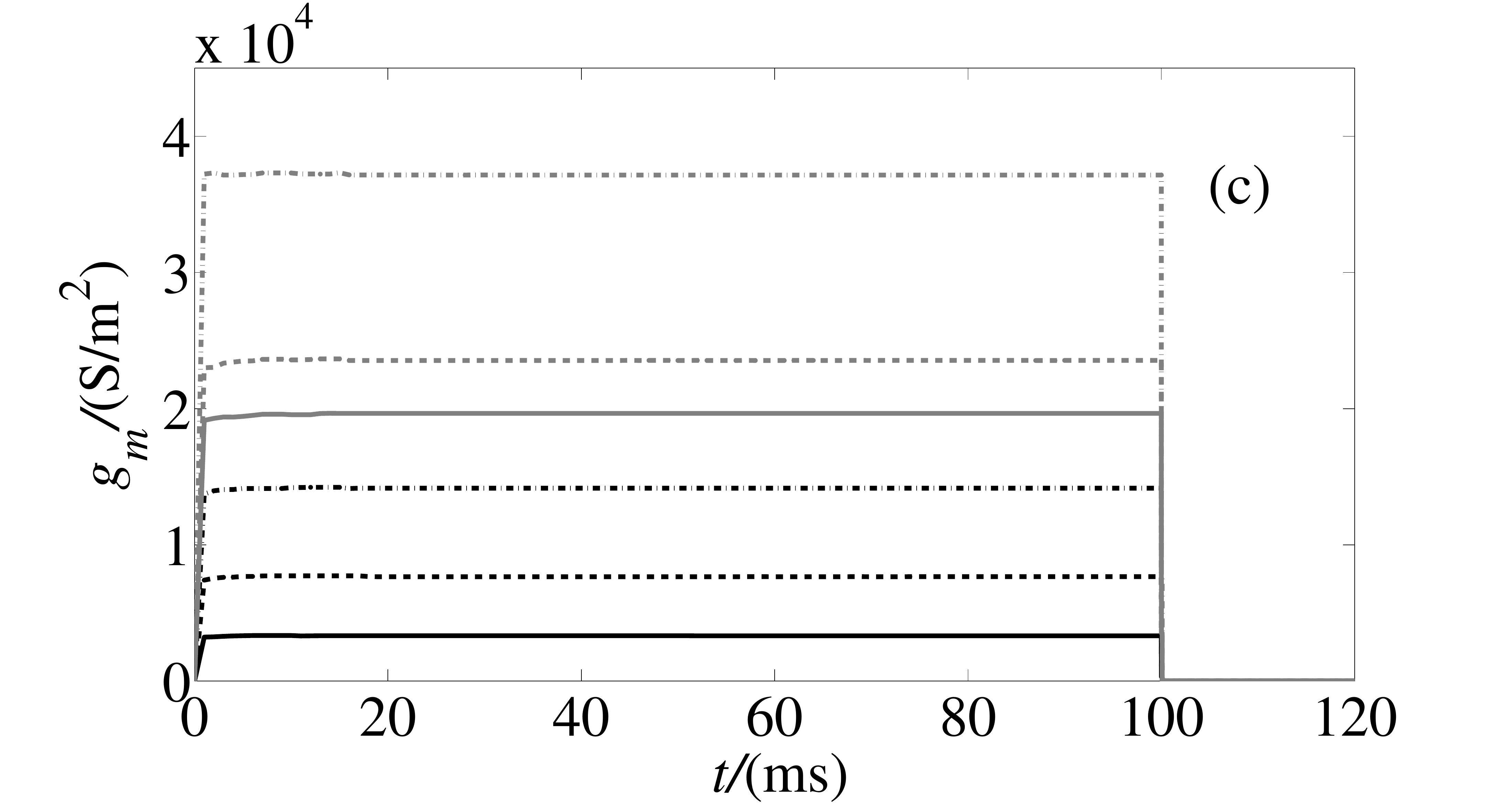}\includegraphics[width=0.5\linewidth]{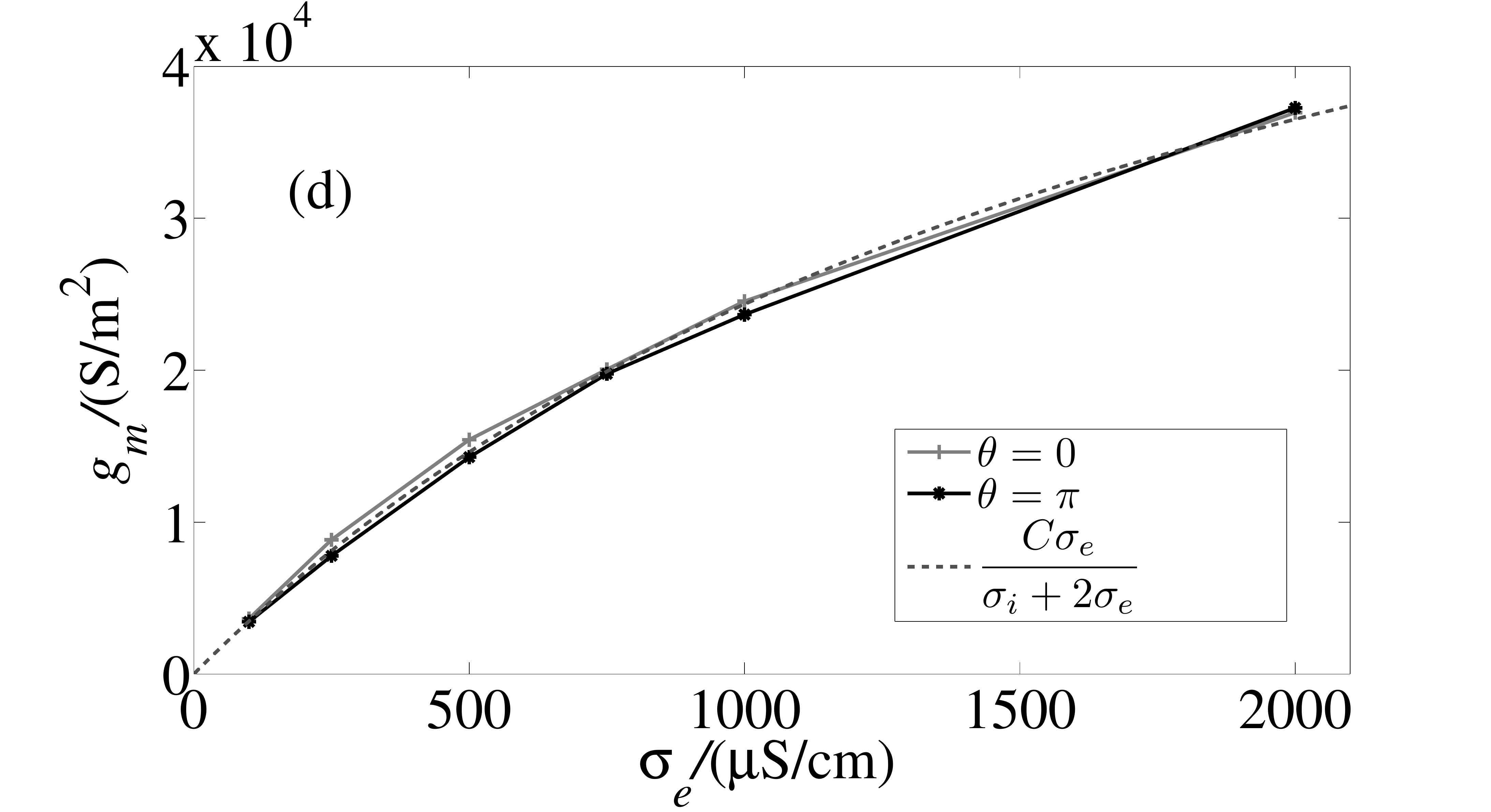}\\\includegraphics[width=0.5\linewidth]{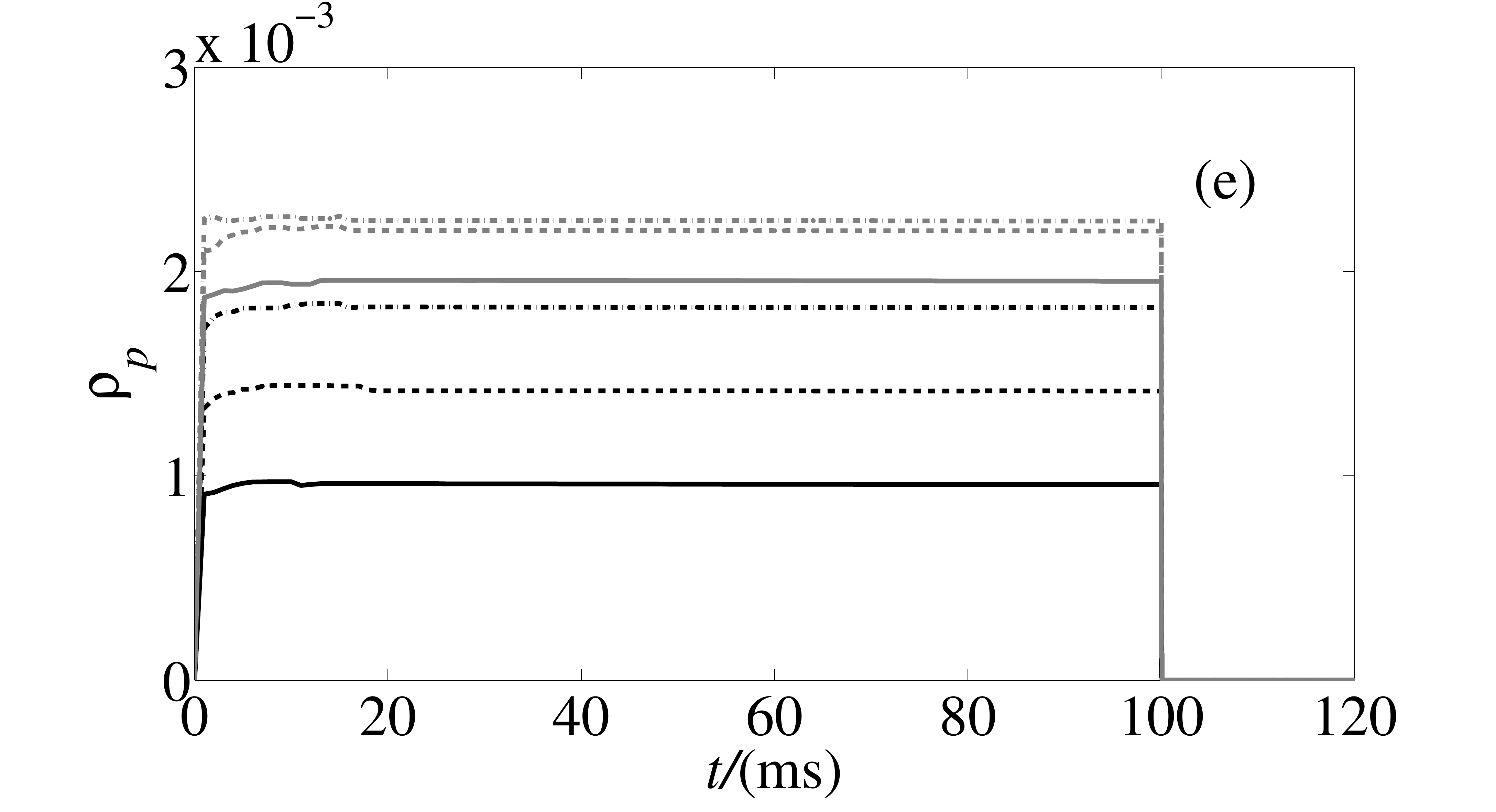}\includegraphics[width=0.5\linewidth]{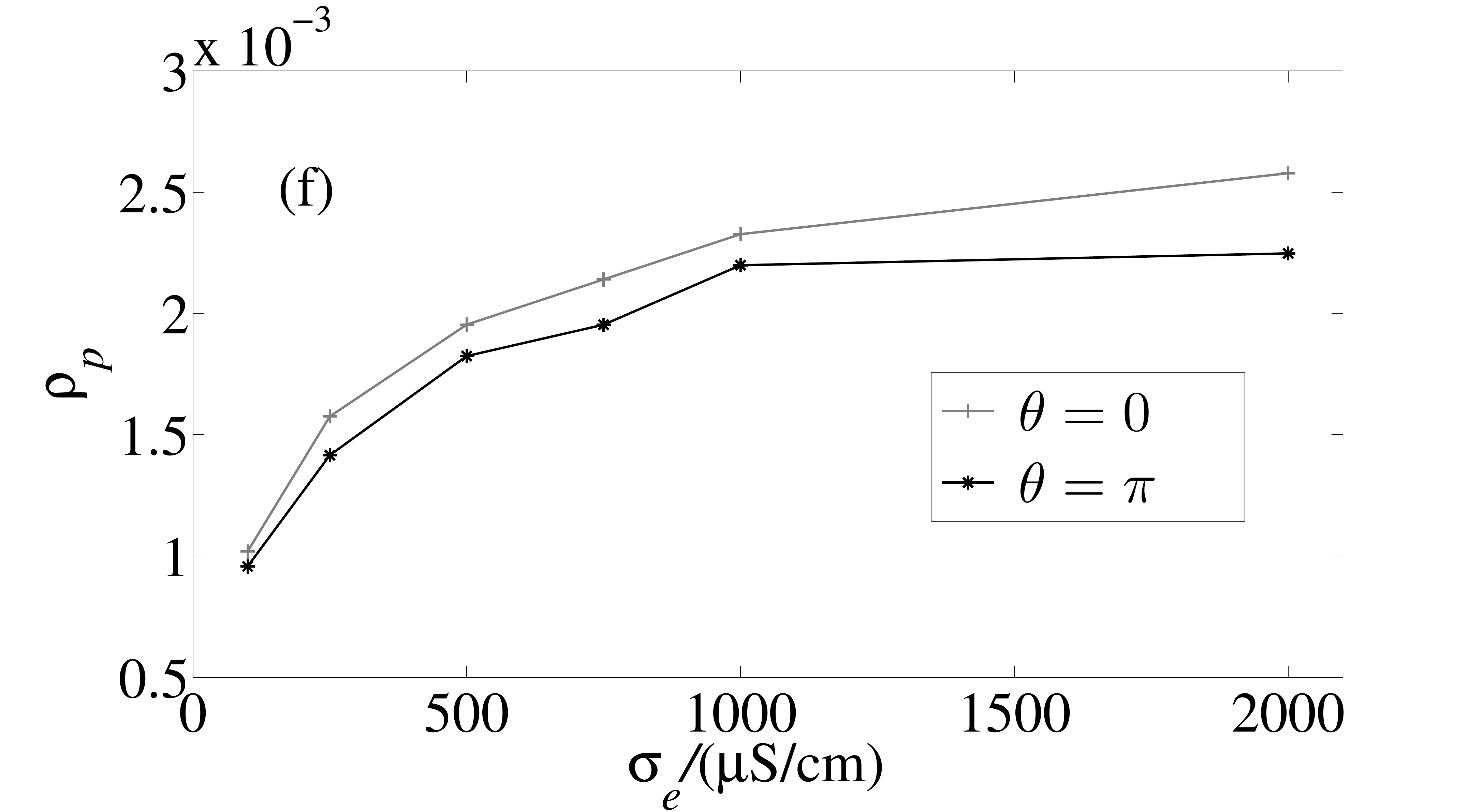}

\caption{Simulated membrane permeabilization under a single pulse of 0.8 kV/cm
in strength and 100 ms in length, for various extra-cellular conductivities.
(a) Evolution of the TMP, $V_{m}$, at $\theta=\pi$ as a function
of time. (b) Polar distribution of the TMP at $t=95$ ms. (c) Evolution
of local membrane conductance, $g_{m}$, at $\theta=\pi$ as a function
of time. (d) Membrane conductance at $\theta=0,\,\pi$ at $t=95$
ms. The dashed is a theoretical prediction, and the fitting constant
$C=1.46\times10^{5}\;{\rm S/m^{2}}$. (e) Evolution of the PAD, $\rho_{p}$,
at $\theta=\pi$ as a function of time. (f) The PAD at $\theta=0,\,\pi$
at $t=95$ ms. \label{fig:VmGmRhop}}
\end{figure}

Figure \ref{fig:VmGmRhop} summarizes the results on the TMP, $V_{m}$,
the effective membrane conductance, $g_{m}$, and the PAD, $\rho_{p}$. Figure \ref{fig:VmGmRhop}a
shows the evolution of $V_{m}$ at $\theta=\pi$ as a function of
time. The differences between the cases are only visible in the initial
stage ($\sim$10 $\mu$s, see the inset), which is caused by the dependence
of charging time on the extra-cellular conductivity \cite{Vlahovska_2009,Sadik_2011}.
For $t>10$ $\mu$s, $V_{m}$ settles to an equilibrium value, which
is maintained until the end of the pulse. Figure \ref{fig:VmGmRhop}b
shows $V_{m}$ as a function of the polar angle, $\theta$, at $t=95$
ms. The consistency of the equilibrium distribution with respect to
the extra-cellular conductivity is evident. We find that this equilibrium
value of $V_{m}$ in the permeabilized regions is determined by a
critical point of the pitchfork bifurcation in the $(r_{eq},V_{m})$
space, where $r_{eq}$ is the equilibrium pore size at a given voltage.
In other words, it is determined by the energy landscape of the porated
membrane, which does not change with respect to the extra-cellular
conductivity. A more detailed analysis is presented in Appendix B
for interested readers. 

Figure \ref{fig:VmGmRhop}c shows the evolution of $g_{m}$ as a function
of time. Similar to $V_{m}$, it exhibits an initial stage of rapid
growth, followed by a plateau (an equilibrium) in the presence of
the pulse, and a rapid decay post-pulsation. However, the equilibrium
value depends strongly and positively on $\sigma_{e}$. This trend
is more obviously observed in Fig. 2d, where $g_{m}$ at $\theta=0,\:\pi$
and $t=95$ ms is plotted against $\sigma_{e}$. This correlation
is derived from the global Ohmic current balance. In fact, following
an analysis similar to that presented in \cite{Jianbo_2012}, we can
show
\begin{equation}
g_{m}\propto\frac{\sigma_{e}}{2\sigma_{e}+\sigma_{i}}(3-\frac{2}{E_{0}a}{\rm max}(|V_{m}|)).\label{eq:gm}
\end{equation}
The details are not presented here for brevity. In Fig. \ref{fig:VmGmRhop}d,
the dashed line represents a fitting in the form $C\sigma_{e}/(2\sigma_{e}+\sigma_{i})$,
where the fitting constant $C=1.46\times10^{5}$ ${\rm S/m^{2}}$.
This result is in qualitative agreement with the numerical study by
Suzuki et al. \cite{Suzuki_2011}. 

Figures \ref{fig:VmGmRhop}e and f show the behavior of $\rho_{p}$
with respect to time and $\sigma_{e}$. Not surprisingly, the trend
concurs with that of $g_{m}$, as the change in the latter is only
caused by a change in membrane permeabilization. Note that this result
qualitatively defers from that in supra-electroporation, where our
model did not indicate a strong dependence of $\rho_{p}$ on $\sigma_{e}$
\cite{Jianbo_2012}. However, in neither situation does membrane permeabilization
provide a viable explanation for the negative correlation between
delivery and extra-cellular conductivity {\cite{Sadik13,Djuzenova_1996,Muller_2001}},
and alternative mechanisms need to be identified. 

\begin{figure}
\includegraphics[width=1\linewidth]{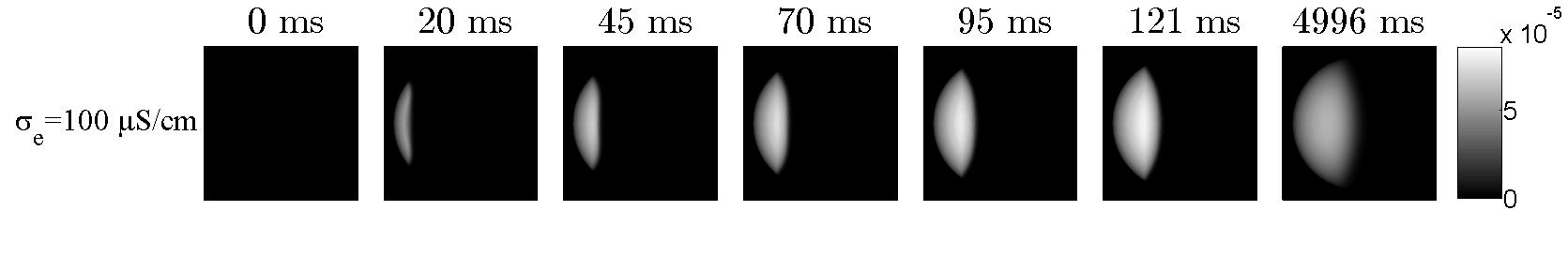}

\includegraphics[width=1\linewidth]{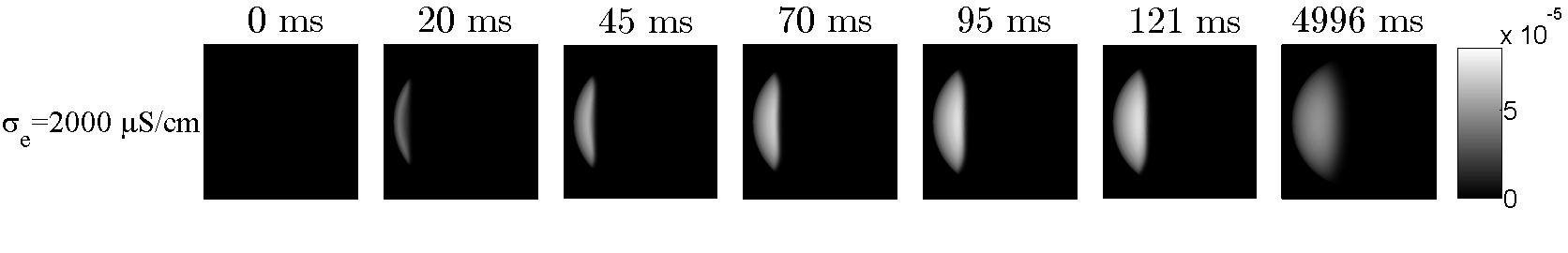}\caption{\label{fig:contour}Contour plots of convoluted PIB concentration at
the cell center-plane for $\sigma_{e}=100\;\mu$S/cm (top) and $2000\; u$S/cm
(bottom). The snapshots are taken at $t=$0, 20, 45, 70, 95, 121,
and 4996 ms, following the experimental presentation in Sadik13.}
\end{figure}

Figure \ref{fig:contour} shows exemplary simulated results in an
attempt to reproduce the experimental fluorescence images in Sadik13
(Fig. 1 therein). The contour plot is based on the convoluted concentration
of PIB:
\begin{equation}
{\rm [PIB]_{conv}}=\int_{-\sigma_{z}/2}^{\sigma_{z}/2}{\rm [PIB]}e^{-z^{2}/2\sigma_{z}^{2}}\mathrm{d{\it z}},\label{eq:conv}
\end{equation}
where $\sigma_{z}$ is the focal depth of the microscopic system,
and $z$ is the axis perpendicular to image acquisition. This convolution
is taken to approximate the effects of a finite focal depth in the
experimental measurements \cite{Jianbo_2011}. The evolution with
respect to the lowest (100 $\mu$S/cm) and the highest (2000 $\mu$S/cm)
conductivities is shown, which is in qualitative agreement with data.
Noticeably, the spread is stronger in case of $\sigma_{e}=100$ $\mu$S/cm.

\begin{figure}
\includegraphics[width=0.5\linewidth]{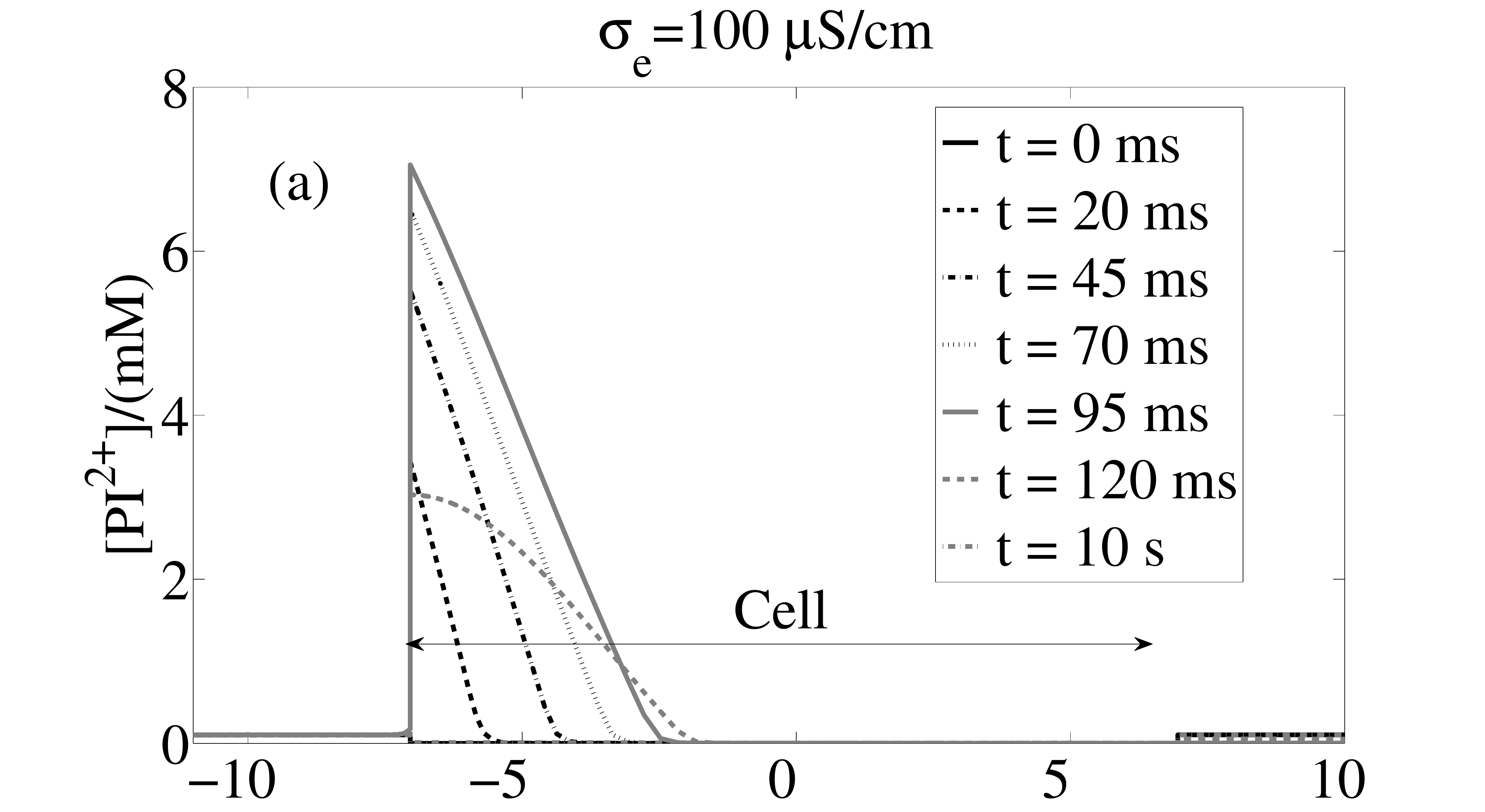}\includegraphics[width=0.5\linewidth]{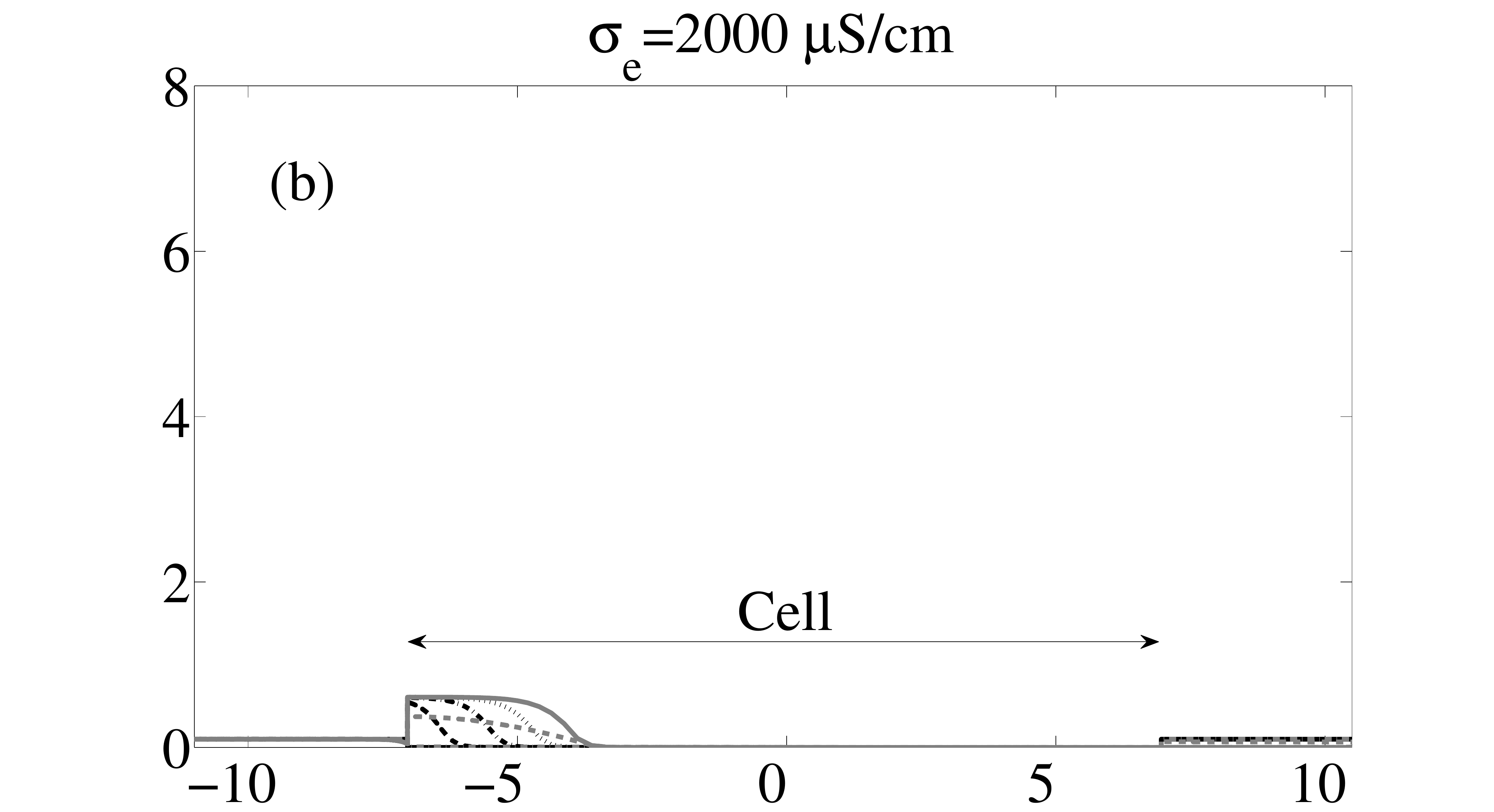}\\\includegraphics[width=0.5\linewidth]{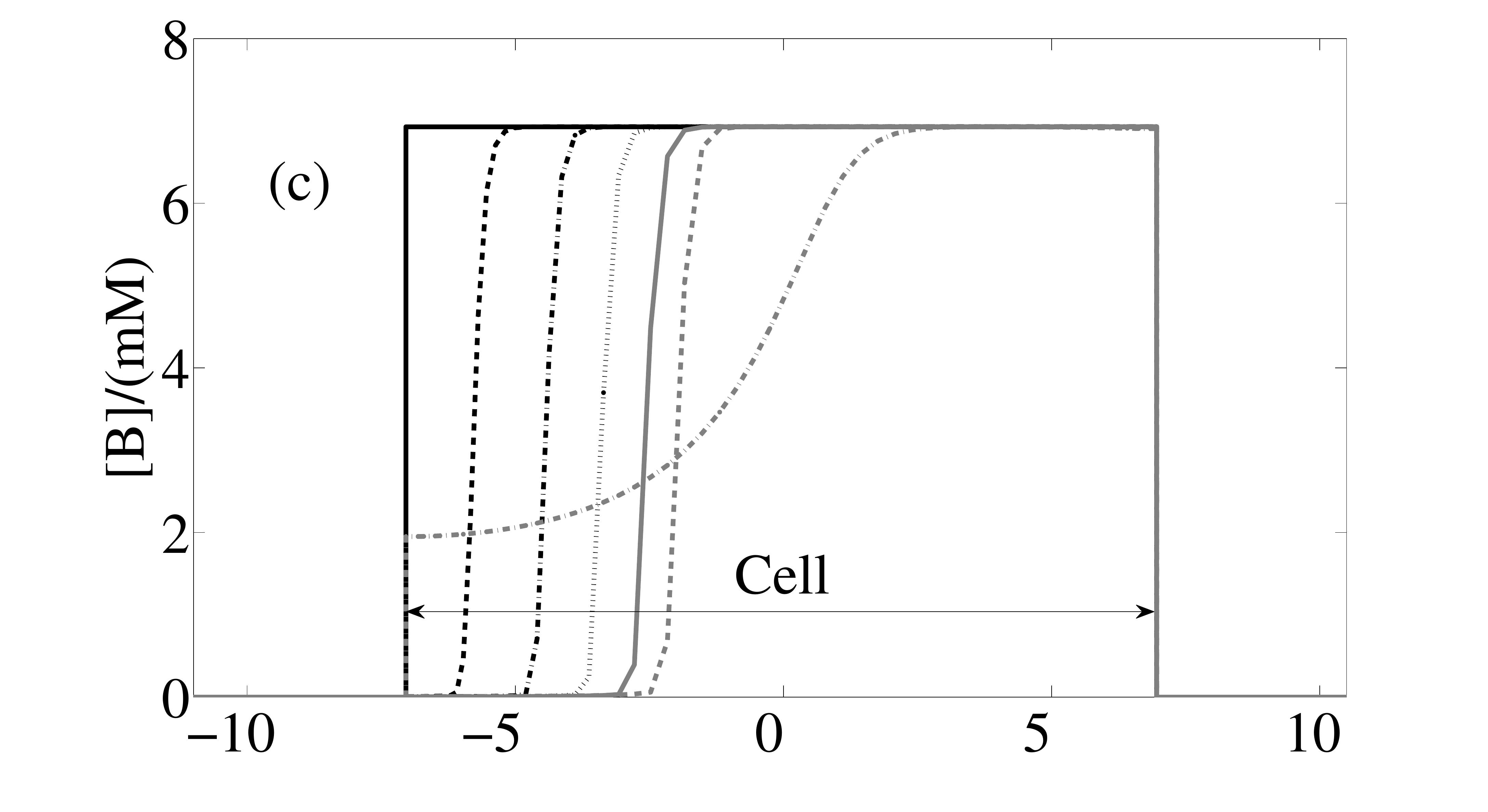}\includegraphics[width=0.5\linewidth]{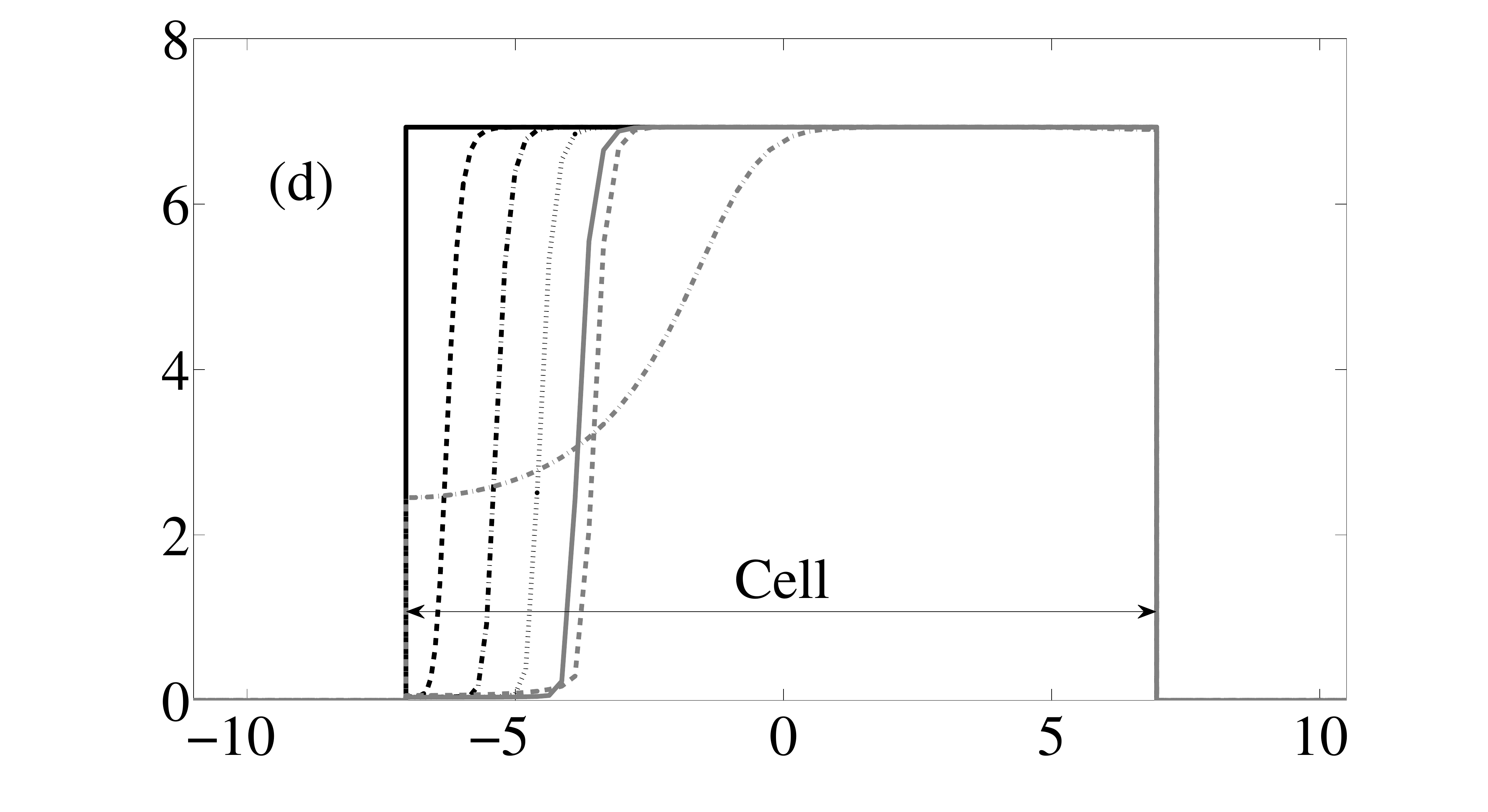}\\\includegraphics[width=0.5\linewidth]{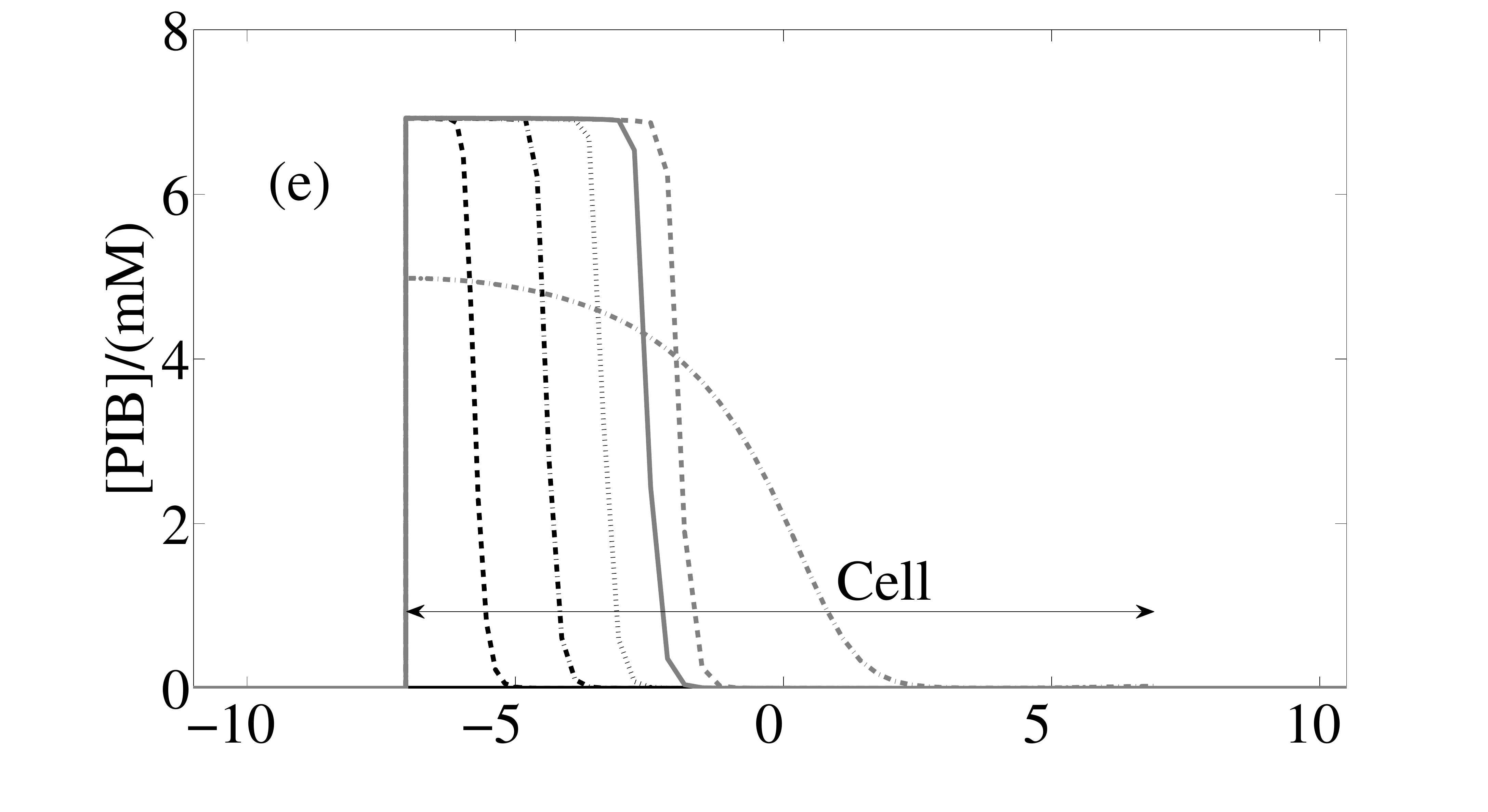}\includegraphics[width=0.5\linewidth]{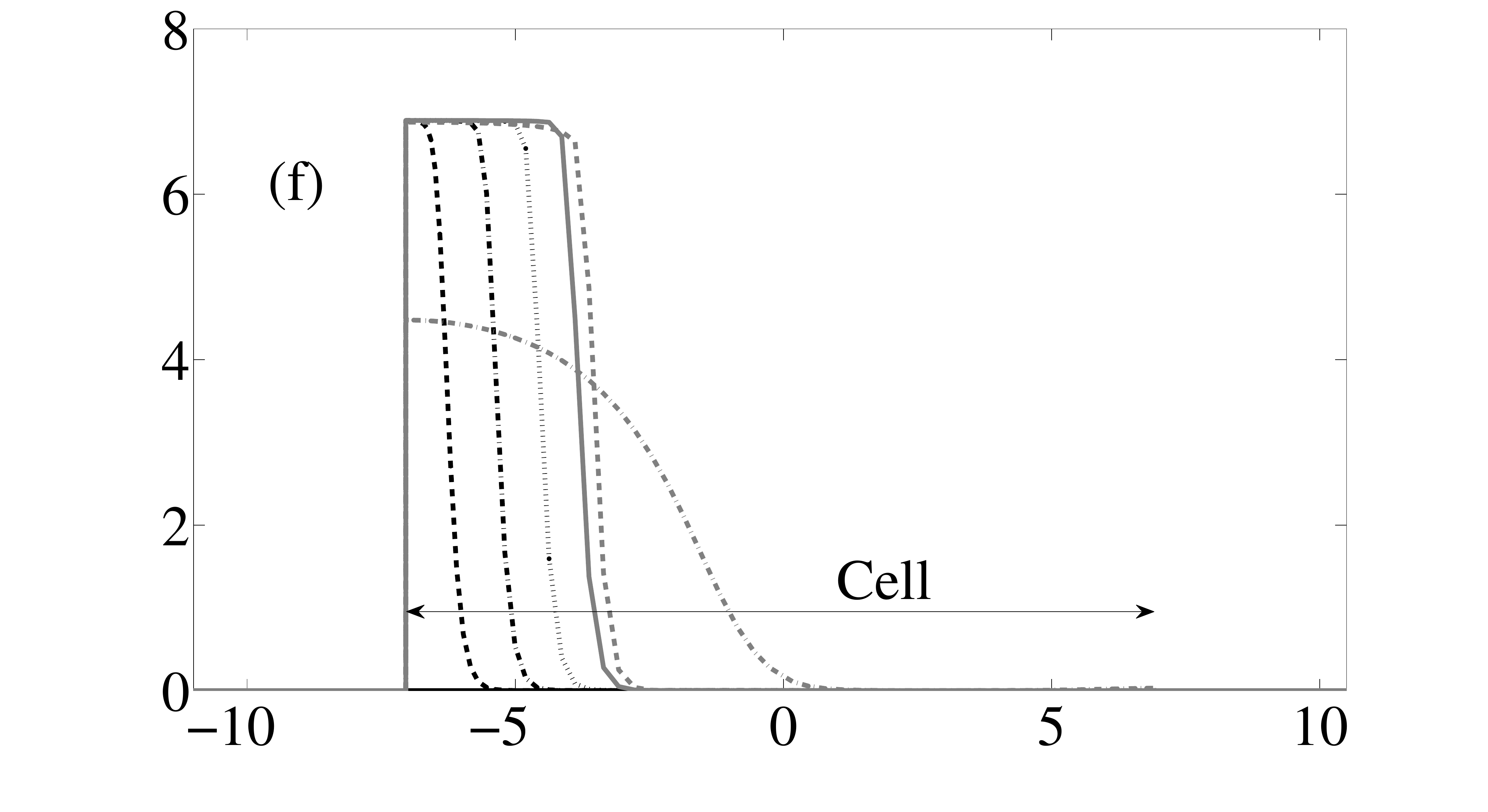}\\

\includegraphics[width=0.5\linewidth]{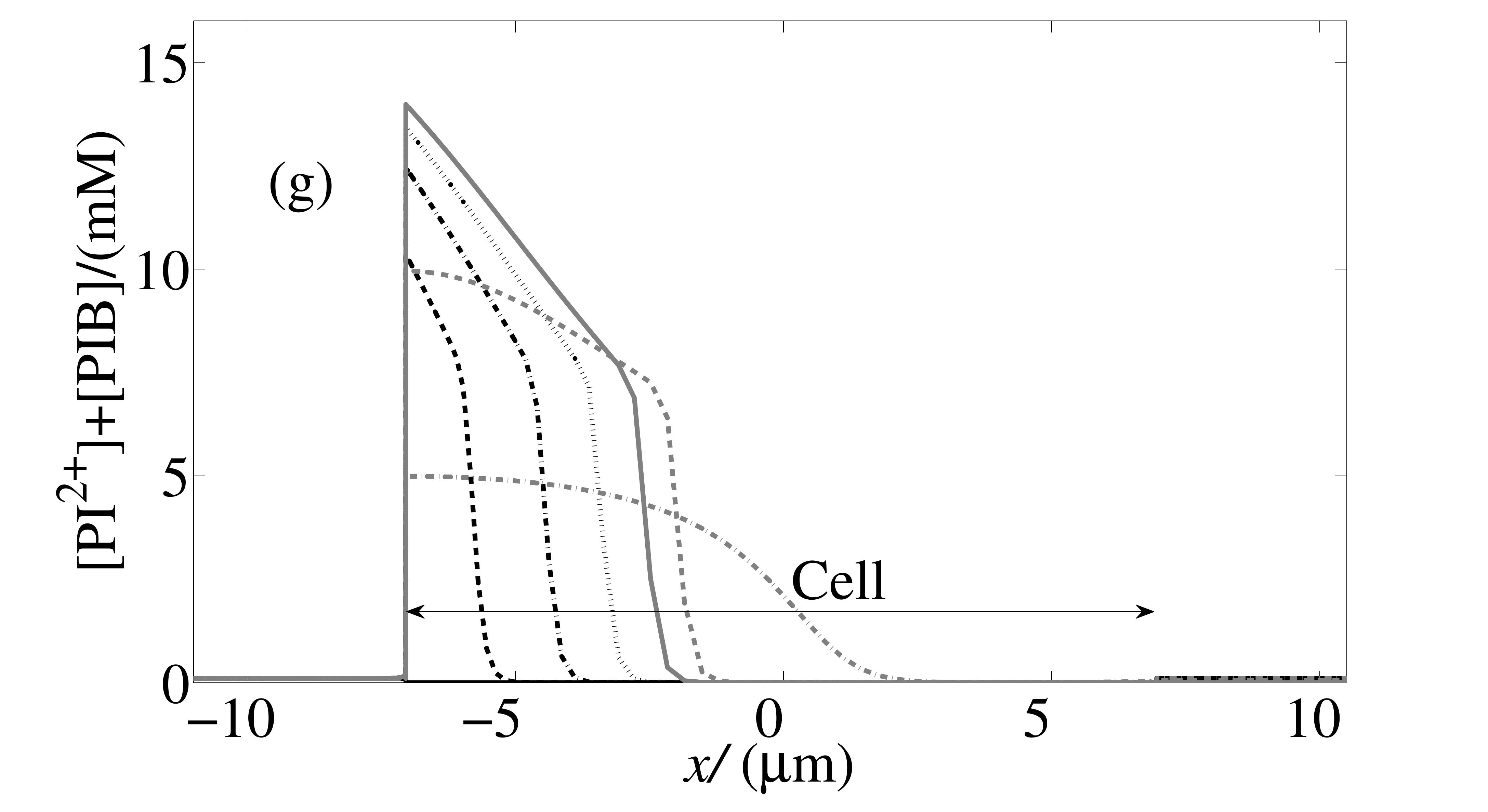}\includegraphics[width=0.5\linewidth]{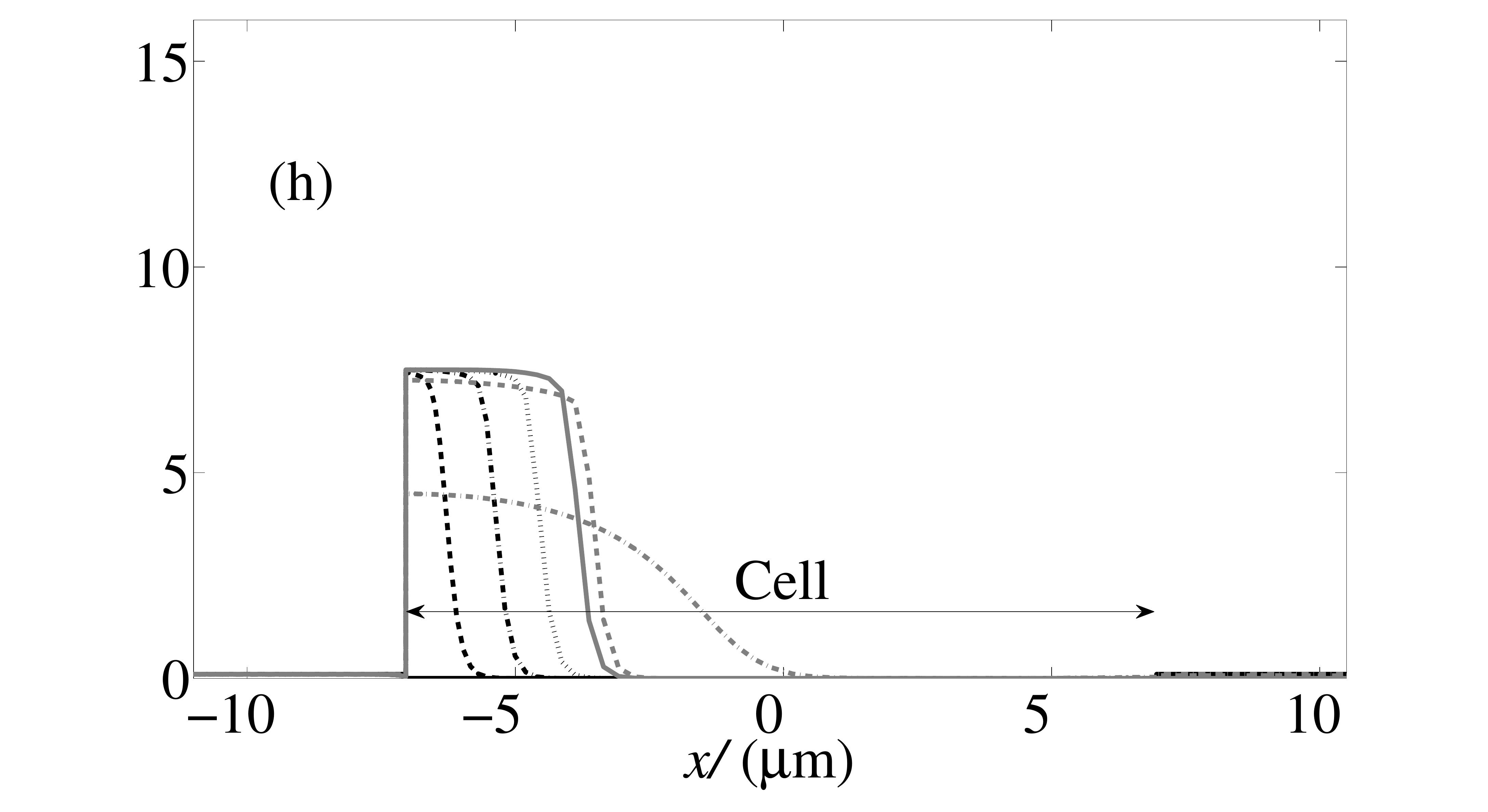}

\caption{Species concentration evolution for $\sigma_{e}=$ 100 $\mu$S/cm
and 2000 $\mu$S/cm along the cell centerline, $x$.\label{fig:centerline}}
\end{figure}

Figure \ref{fig:centerline} demonstrates the detailed evolution of
the species concentrations, also for the two extreme values of $\sigma_{e}$.
Figures \ref{fig:centerline}a-d show $\mbox{[\mbox{P\ensuremath{\mbox{I}^{2+}}}]}$
and $[\mbox{B}]$ at different times along the cell centerline, $x$.
Together, the results indicate that the binding sites are exhausted
upon electrophoretic entry of the free ions in the presence of the
100-ms pulse. Continuous intra-cellular diffusion and association/dissociation
occur after the pulse ceases. However, no appreciable molecular exchange
across the membrane is predicted. Figures \ref{fig:centerline}e and
f show the {[}PIB{]} profile. This compound is responsible for the
experimentally observed fluorescence emission, and the convoluted
concentration as defined by Eq. (\ref{eq:conv}) is assumed to be
proportional to the fluorescence intensity. During the pulse, the
front of {[}PIB{]} profile advances uniformly along the field direction
due to binding-site exhaustion. The redistribution post-pulsation
is due to intra-cellular redistribution of all species. In comparison
with Fig. \ref{fig:contour}, the peaks observed therein are attributed
to the convolution over a spherical cell geometry, which effect we
have previously explained \cite{Jianbo_2011}. Finally, Figs. \ref{fig:centerline}g
and h show the sum of the free and bound ions. This quantity indicates
the total PI concentration in the cell. For both values of $\sigma_{e}$,
this summed concentration ($\sim10$ mM) is significantly higher than
the extra-cellular PI concentration (100 $\mu$M). Furthermore, consistent
with the experimental observation, delivery decreases when $\sigma_{e}$
increases. These trends are explained with an electrokinetic phenomenon
termed Field-Amplified Sample Stacking (FASS), which is discussed
in greater details in our previous work \cite{Jianbo_2011,Jianbo_2012}.

\begin{figure}[t]
\includegraphics[width=0.5\linewidth]{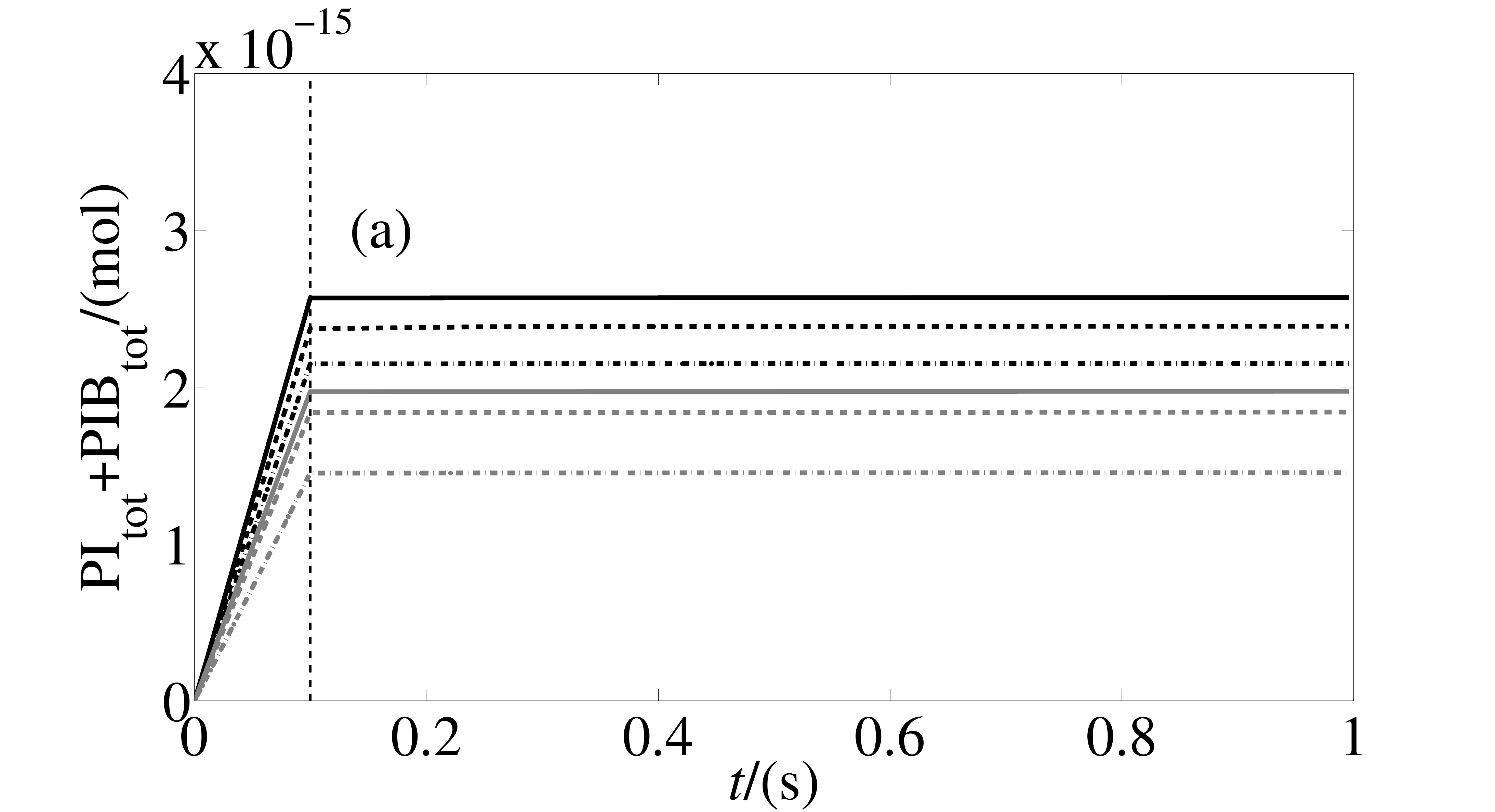}\includegraphics[width=0.5\linewidth]{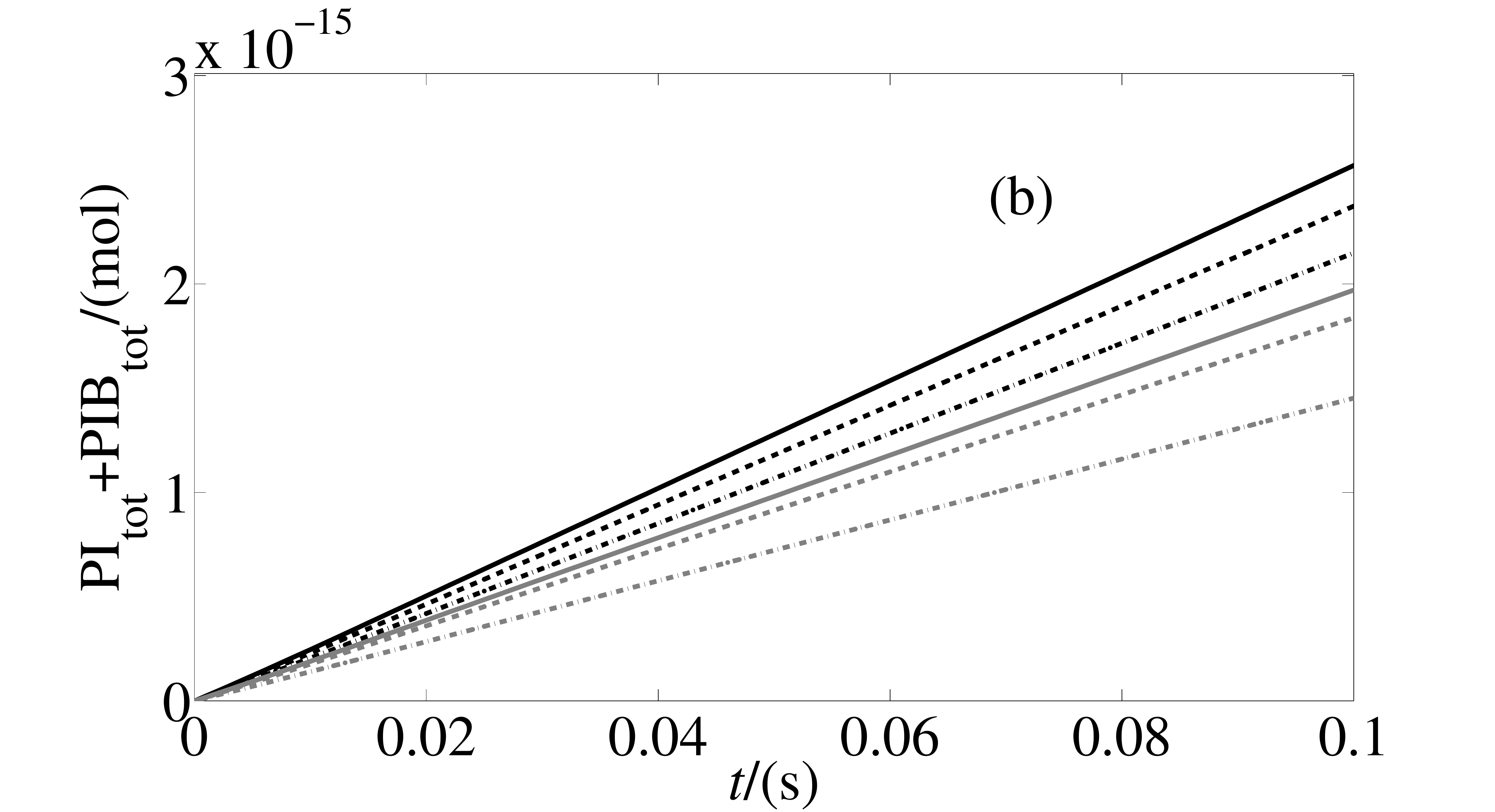}\\\includegraphics[width=0.5\linewidth]{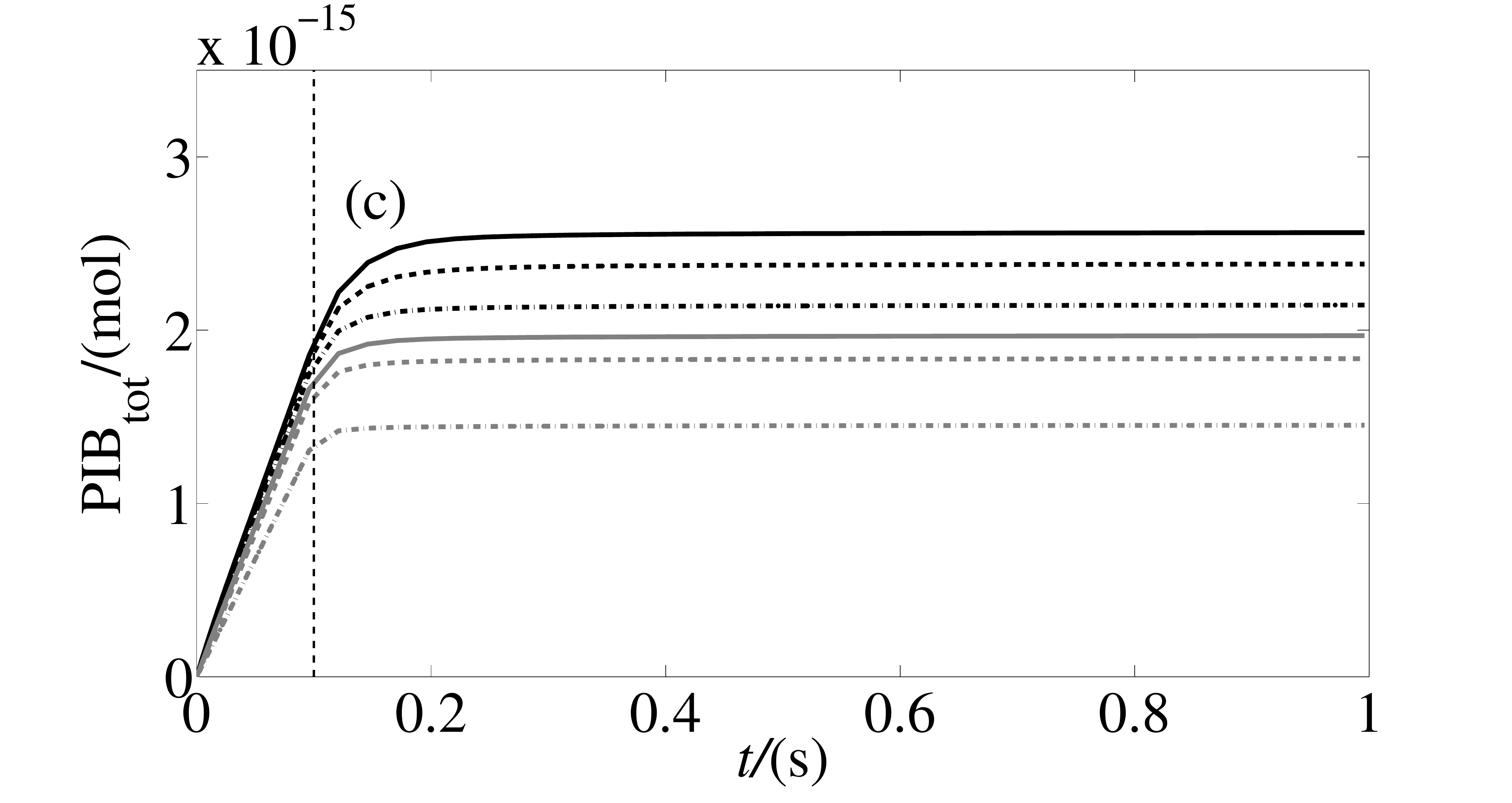}\includegraphics[width=0.5\linewidth]{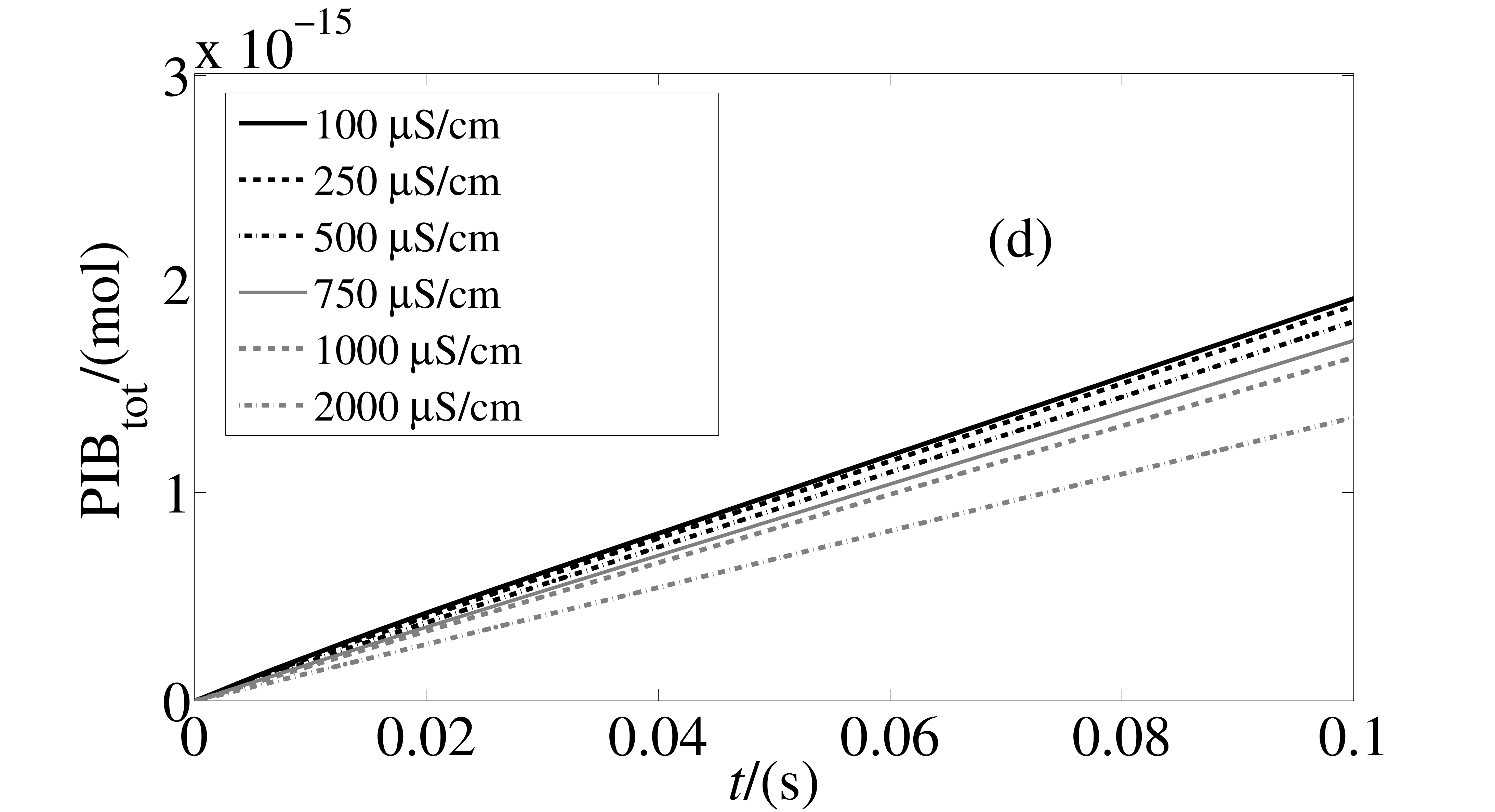}\caption{(a) Evolution of integrated free PI and PIB concentrations over the
entire cell (${\rm PI_{tot}+PIB_{tot}}$) for the six values of $\sigma_{e}$.
The dashed vertical line indicates the end of the pulse. (b) Evolution
of ${\rm PI_{tot}+PIB_{tot}}$ during the pulse. (c) Evolution of
integrated PIB concentration over the entire cell (${\rm PIB_{tot}}$).
(d) Evolution of ${\rm PIB_{tot}}$ during the pulse. \label{fig:evolution}}
\end{figure}

\begin{figure}[t]
\includegraphics[width=0.5\linewidth]{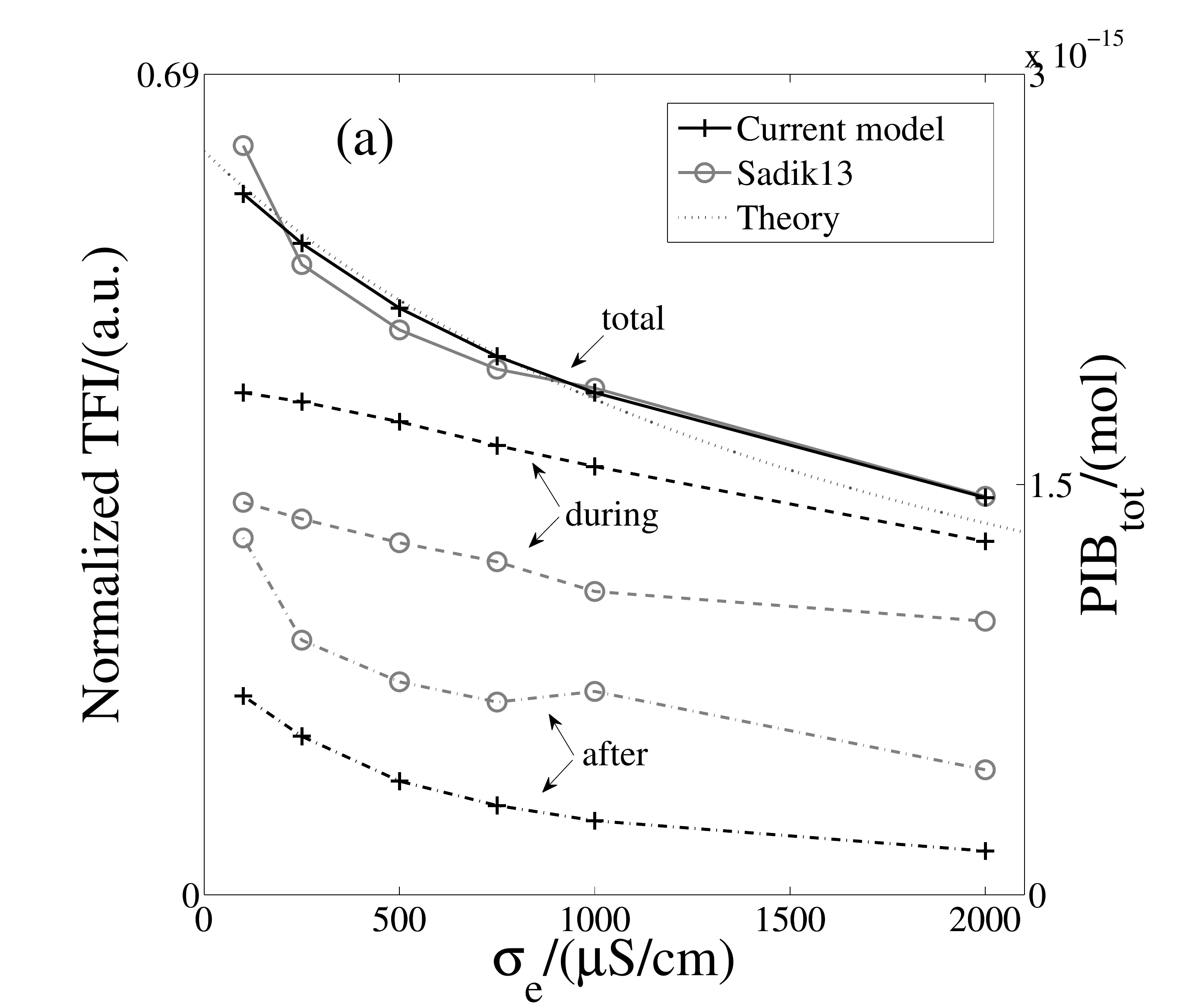}\includegraphics[width=0.5\linewidth]{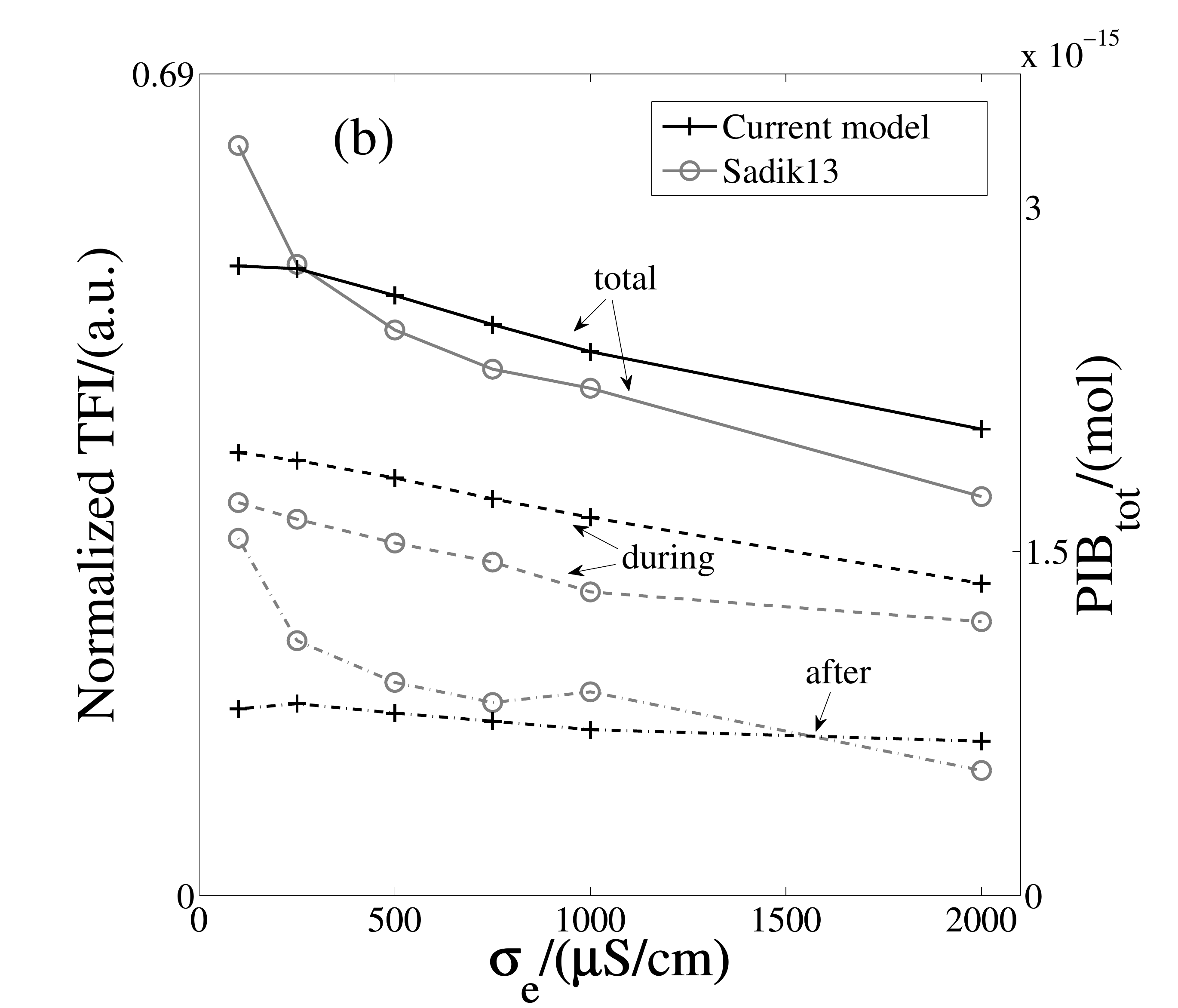}\caption{Comparison of simulated results ($'+'$, right axis) with experimental
data from Sadik13 ($'{\rm o}'$, left axis). For the data, the middle
and lower curves represent contributions to the total fluorescence
intensity (TFI) during and after the pulse, respectively. The upper
curve is the sum of the two. In comparison, numerical simulation of
${\rm PIB_{tot}}$ is presented, and the upper, middle, and lower
curves are defined similarly. The right axis is scaled linearly such
that the upper curves from data and simulation best match. (a) Simulated
results using the original ASE model. The solid line is a theoretical
fitting with the functional form $C/(2\sigma_{e}+\sigma_{i})$. (b)
The size of the pores are artificially maintained for 1 s post-pulsation
in the ASE model to allow for more diffusive delivery.\label{fig:PIBtot}}
\end{figure}

To study total delivery, we integrate and sum the free and bound PI
concentrations over the entire cell, and denote the resulting quantity
by ${\rm PI_{tot}+PIB_{tot}}$. The evolution of total delivery as
a function of time and for the six values of $\sigma_{e}$ is shown
in Figs. \ref{fig:evolution}a and b, where Fig. \ref{fig:evolution}b
displays the specific stage during the pulse. Once the pulse ceases,
the total delivery, ${\rm PI_{tot}+PIB_{tot}}$, does not further
increase. This saturation of delivery is due to the fact that in the
model, pores immediately return to a very small size ($r_{m}$= 0.8
nm, \cite{Krassowska2007}), hence significantly diminishing $\rho_{p}$
and preventing post-pulsation diffusive delivery. A further examination
on the effect of the latter is presented later in Fig. \ref{fig:PIBtot}b.
Figures \ref{fig:evolution}c and d show the evolution of bound PI
integrated over the whole cell (${\rm PIB_{tot}}$). The results are
in qualitative agreement with the total fluorescence intensity (TFI,
induced by the compound PIB) presented in Sadik13. Figure \ref{fig:evolution}c
shows, in contrast to Fig. \ref{fig:evolution}a, that ${\rm PIB_{tot}}$
continues to increase even after the pulse ceases. Although no more
PI is delivered into the cell at this stage, the available free PI
ions that have already entered the cell during the pulse (Figs. \ref{fig:centerline}a
and b) continue to spread and bind, causing ${\rm PIB_{tot}}$ to
further increase. A final steady-state is reached within the diffusive
time scale ($\sim$ 0.1 s) after the association/dissociation processes
equilibrate over the entire cell. 

Together, Figs. \ref{fig:centerline} and \ref{fig:evolution} impart
important insights. First, due to the FASS mechanism and the high
conductivity ratio, the cell can be ``loaded'' with a high concentration
of ions via electrophoretic transport, even with short pulses. In
the current simulations, this ``loading'' is sufficient to exhaust
locally the high concentration of binding sites. A similar conclusion
can be drawn if PI is replaced by other target agents such as drug
molecules. Second, the observed increase in fluorescence signal post-pulsation,
such as that presented in Sadik13 may be partially attributed to this
``pre-loading'' effect. Therefore, caution needs to be taken to
interpret experimental data where an indirect indicator such as PIB
is used to study PI delivery. 

Figure \ref{fig:PIBtot}a compares the simulated results with data
from Sadik13. The experimental data is denoted by circles, and the
middle and lower curves represent contributions to the normalized
TFI during and after the pulse, respectively. The upper curve is the
sum of the two. The model prediction is denoted by pluses, and the
definition of the curves follows the data. Because of the difference
in the units of the measurement (a.u. for the fluorescence signal)
and the prediction (mol for ${\rm PIB_{tot}}$), the axis presenting
the latter is scaled linearly, such that the upper curves are best
matched. The comparison demonstrates that the trends in the data are
very-well captured by the numerical study, although we observe quantitative
differences between the middle and lower curves. As noted above, in
the model prediction, although ${\rm PIB_{tot}}$ continues to increase
after the pulse, it is in actuality attributed to further binding
of the free PI ions already delivered into the cell during the pulse.

The correlation between ${\rm PIB_{tot}}$ and $\sigma_{e}$ can be
approximated by
\begin{equation}
{\rm PIB_{tot}\propto\frac{1}{2\sigma_{e}+\sigma_{i}}}.\label{eq:PIBtot}
\end{equation}
A fitting using this functional form is shown as the dotted line in
Fig. \ref{fig:PIBtot}a. This correlation can be derived if we assume
that delivery is primarily mediated by electrophoresis. The molar
flux into the cell is proportional to $c_{e}E_{e}$, where $c_{e}$
is the extra-cellular concentration of ${\rm PI^{2+}}$, and is assumed
to be constant (${\rm [PI^{2+}]}_{e,o}$). Therefore, molecular delivery
is proportional to the extra-cellular field strength at the membrane.
Using Eq. (\ref{eq:Ohmic}), the steady-state expression of $E_{e}$
is given by 
\begin{equation}
E_{e}=\frac{g_{m}V_{m}}{\sigma_{e}}.\label{eq:Ee}
\end{equation}
Substituting Eq. (\ref{eq:gm}) into Eq. (\ref{eq:Ee}), and considering
$V_{m}$ does not change with respect to the extra-cellular conductivity,
we arrive at the correlation (\ref{eq:PIBtot}). 

In the simulation above, and similar to our previous studies \cite{Jianbo_2011,Jianbo_2012},
$\rho_{p}$ decreases by three orders of magnitude immediately after
the pulse ceases, due to the shrinking of the pore size in the absence
of $V_{m}$. This reduction prohibits appreciable diffusive transport
afterwards. This behavior is an artifact of the ASE model due to its
incapability to include the resealing process (typically on the order
of seconds to minutes \cite{Teissie_2009,Weaver_1996,He_2008}). To
investigate the effects of post-pulsation diffusion, we artificially
prevent $\rho_{p}$ reduction at the end of the pulse. In other words,
we keep $\rho_{p}$ at its value at the end of the pulse for an additional
second. The result is shown in Fig. \ref{fig:PIBtot}b, also in comparison
with the data. Similarly, the axis for ${\rm PIB_{tot}}$ is rescaled
to best match the upper curves. We observe that the respective contributions
from during and after the pulse match better quantitatively when comparing
with Fig. \ref{fig:PIBtot}a. However, the inverse trend with respect
to $\sigma_{e}$ is abated. This result is not surprising, as diffusive
transport correlates positively with $\rho_{p}$, which in turn depends
positively on $\sigma_{e}$ (Fig. \ref{fig:VmGmRhop}f). The addition
of the diffusive delivery therefore weakens the inverse trend observed
in the data. 

In summary, the model prediction agrees qualitatively with the experimental
data in general, and quantitatively in terms of the correlation between
delivery and the extra-cellular conductivity (Eq. (\ref{eq:PIBtot})).
The results confirm that this inverse correlation is primarily mediated
by electrophoretic transport during the pulse. In fact, this trend
tends to be abated rather than enhanced by diffusive transport due
to the positive dependence of permeabilization on extra-cellular conductivity.
The current study suggests that electrophoretic transport may be important
even for a small molecule such as PI.

\section{Conclusions}

In this work, we have implemented a companion model study for the
experimental counterpart by Sadik et al. \cite{Sadik13}. Results
on both membrane permeabilization and molecular delivery are presented,
through which we gather useful insights on the system behavior. 

The TMP in the permeabilized regions exhibits a consistent value across
all six extra-cellular conductivities examined. Through a detailed
investigation, we find that this value corresponds to a bifurcation
point in the relation between equilibrium pore size and the TMP. This
finding bears significance in that it connects the mesoscopic ASE
model with macroscopic observables. In other words, this critical
value was previously specified empirically in the ASE model; with
the current theory, it can be directly measured by fluorescence techniques
following Kinosita et al. \cite{Kinosita_1988} or Flickinger et al.
\cite{Flickinger_2010}. 

Both the PAD and the membrane conductance are predicted to increase
with an increasing extra-cellular conductivity. These correlations
naturally result from the requirement to satisfy the Ohmic current
conservation condition. In fact, the relation between membrane conductance
and extra-cellular conductivity follows the functional form of $g_{m}\propto\sigma_{e}/(2\sigma_{e}+\sigma_{i})$,
which can be derived from an idealized model for the electric potential.
This positive correlation between membrane permeabilization and extra-cellular
conductivity rules out pure diffusive transport as a viable interpretation
for the opposite effect of the latter on delivery. 

For PI delivery, the model correctly predicts the inverse dependence
on extra-cellular conductivity. This agreement confirms that this
behavior is primarily mediated by electrophoretic transport during
the pulse. In fact, diffusion tends to abate rather than enhance the
trend. The correlation between delivery and extra-cellular conductivity
is quantitatively captured by the model, namely, ${\rm PIB_{tot}\propto{\it {\rm 1/({\it {\rm 2}\sigma_{e}+\sigma_{i}}{\rm )}}}}$.
The simulation also reveals that an increase in the fluorescence intensity
after the pulse ceases is not necessarily attributed to molecules
entering the cell during this time; it may rise from continuous spreading
and binding of free ions ``loaded'' into the cell in the presence
of the pulse. Together, the results suggest that electrophoretic transport
is important even for a small molecule such as PI.

Last but not least, a direct comparison between experimental data
and model simulation as presented in this work helps establish confidence
in and validate the latter. In the ASE model, a few parameters (such
as $\beta$, $\gamma$, and $F_{max}$, see Eq. (\ref{eq:app1}) in
Appendix B) are specified empirically following previous work. However,
membrane permeabilization (including the PAD and membrane conductance)
is found not to depend critically on the specific values of these
parameters. Instead, the bifurcation point of the TMP, together with
the Ohmic current conservation law strongly regulate the permeabilization
behavior. The model therefore provides robust predictions which are
useful for the study of electroporation-mediated molecular delivery.

\section*{Acknowledgments}

The authors acknowledge funding support from an NSF Award CBET-0747886
with Dr. Henning Winter and Dr. Dimitrios Papavassiliou as contract monitors.

\section*{Appendix A: Mathematical definition of pertinent variables}

\subsection*{Transmembrane Potential (TMP)}

The transmembrane potential, $V_{m}$, at any membrane location, $\theta$,
is defined by the potential jump across the membrane:

\begin{equation}
V_{m}(t,\theta)=[\Phi_{i}(t,\theta)-\Phi_{e}(t,\theta)]\mid_{r=a}.
\end{equation}

\subsection*{Total ionic current density}

The total ionic current density $j_{p}$ is the summed electric current
density over all electropores on a local element, defined by:

\begin{equation}
j_{p}(t,\theta)=\sum_{j=1}^{K(t,\theta)}i_{p}(r_{j}(t,\theta),V_{m})/\Delta A,\label{eq:jp}
\end{equation}
where $\Delta A$ is the area of the local element, $K$ is the total
number of conductive pores, and $i_{p}$ denotes the current through
an individual pore with radius $r_{j}$ and subject to a transmembrane potential
$V_{m}$. $i_{p}$ is given by the formula:

\begin{equation}
i_{p}=\frac{2\pi r_{j}^{2}\sigma_{eff}V_{m}}{\pi r_{j}+2h},\label{eq:ip}
\end{equation}
where $\sigma_{eff}=(\sigma_{e}-\sigma_{i})/\mbox{ln}(\sigma_{e}/\sigma_{i})$
is an effective pore conductivity, and $h$ is the membrane thickness.
A detailed derivation of Eq. (\ref{eq:ip}) can be found in our previous
work \cite{Jianbo_2010}.

\subsection*{Pore area density (PAD)}

The pore area density $\rho_{p}$ is the fractional area occupied
by the conductive pores at a specific point on the membrane, defined by the following formula:

\begin{equation}
\rho_{p}(t,\theta)=\sum_{j=1}^{K(t,\theta)}\pi r_{j}^{2}/\Delta A, \label{eq:rho_p}
\end{equation}
where the definition of $\Delta A$, $K$ and $r_{j}$ are the same
as in Eq. (\ref{eq:jp}). $\rho_{p}$ is used as a measurement of the
membrane permeabilization level. 

\section*{Appendix B: The equilibrium transmembrane potential}

In Figs. \ref{fig:VmGmRhop}a and b, we find that the TMP, $V_{m}$, settles
to an equilibrium value in the permeabilized regions within a few
microseconds after the pulse starts. Furthermore, this value does
not vary appreciably with respect to the extra-cellular conductivity.
Here we argue that this value is determined by a critical point in
the pitchfork bifurcation in the ($r_{eq},V_{m}$) relation, where
$r_{eq}$ is the equilibrium pore size at a given voltage. 

\begin{figure}
\includegraphics[width=0.5\linewidth]{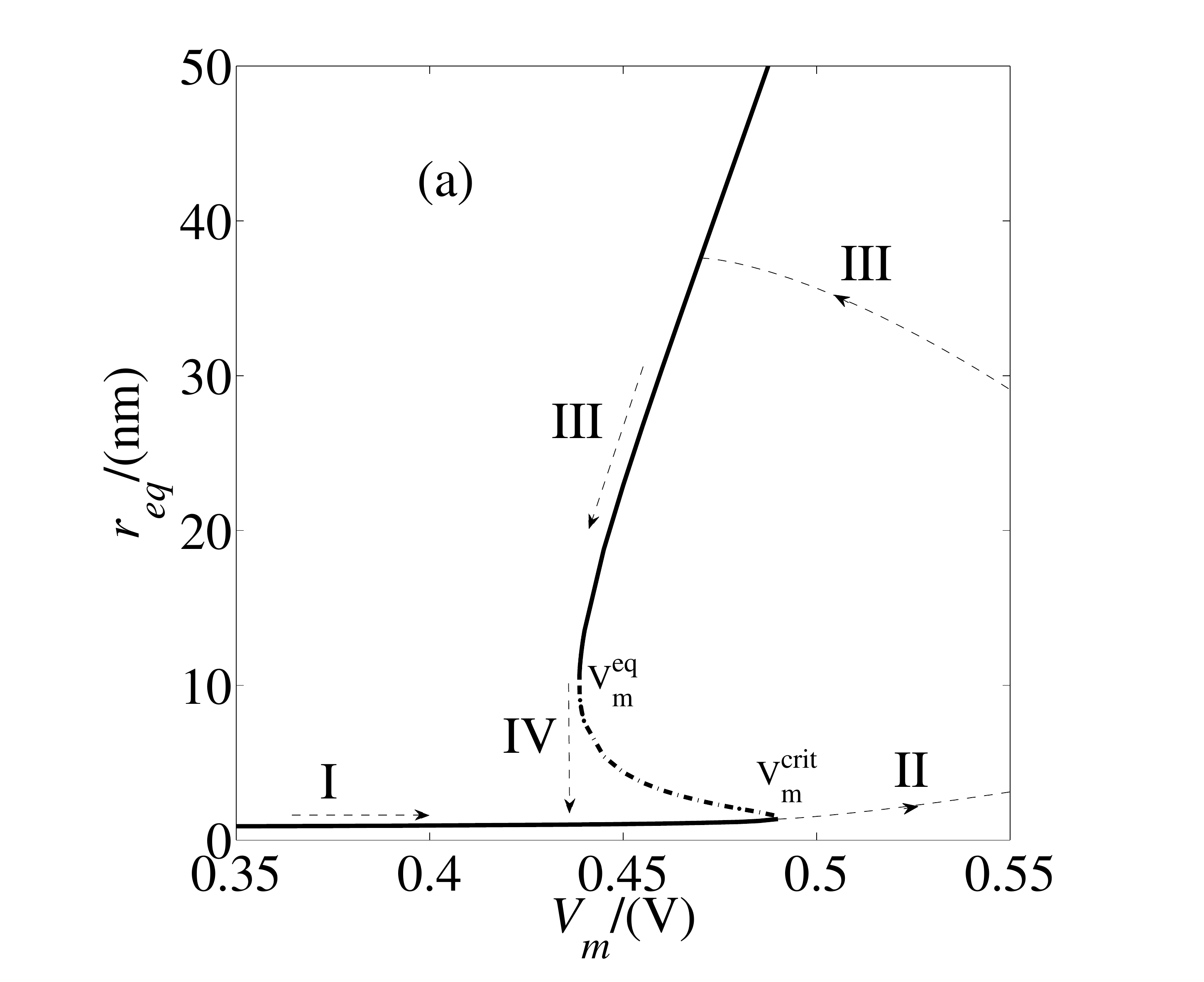}\includegraphics[width=0.5\linewidth]{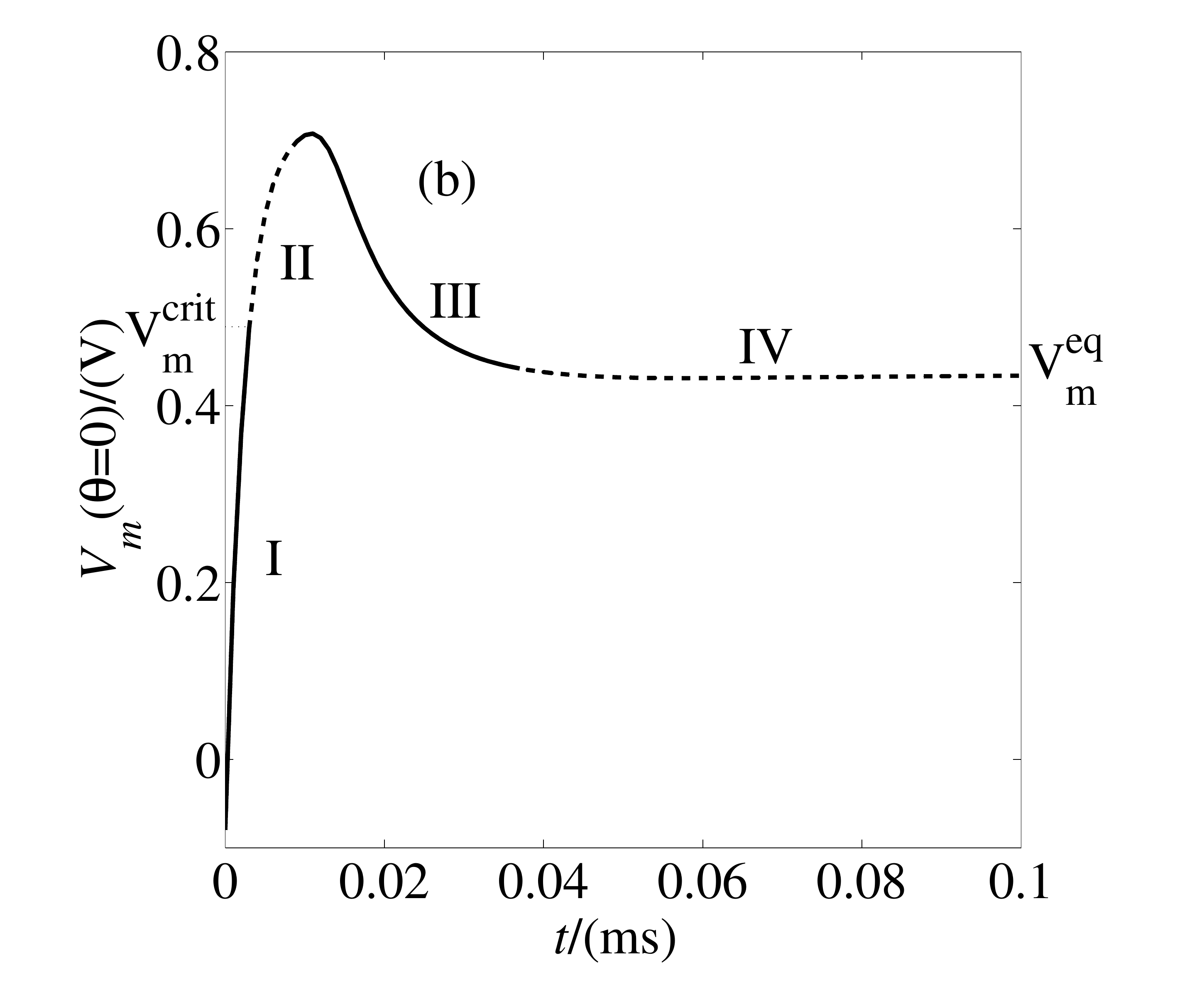}\caption{Schematics for the ($r_{eq},V_{m}$) dynamics. (a) The relation between
$r_{eq}$ and $V_{m}$ exhibits a pitchfork bifurcation behavior.
Between $V_{m}^{eq}$ and $V_{m}^{crit}$, three branches of solutions
exist from Eq. (\ref{eq:req_2}), where the middle one (dash-dotted)
is unstable. The initial charging and permeabilization process follows
four stages denoted by I-IV. For this case $\rho_{p}=2\times10^{-3}$.
A detailed description is given in the text. (b) Exemplary evolution
of $V_{m}$ as a function of time. The stages I-IV correspond to those
in (a). \label{fig:bifurcation}}

\end{figure}

In the ASE we use, and in general in the Smoluchowski equations governing
the pore dynamics \cite{Neu_1999,Krassowska2007,Freeman_1994}, the
pore size evolves to minimize membrane energy. The rate of change
is given by the equation \cite{Krassowska2007}:

\begin{equation}
\dot{r}=\frac{D_p}{K_{B}T}\Biggl[\frac{4r_{p}^{4}\beta}{r^{5}}-2\pi\gamma+2\pi\sigma_{eff}{r}+\frac{F_{max}V_{m}^{2}}{1+\frac{r_{h}}{{r}+r_{t}}}\Biggr],\label{eq:app1}
\end{equation}
where $r$ is the pore radius, $D_p$ is the pore radius diffusion coefficient,
$K_{B}$ is the Boltzmann constant, and $T$ is temperature. The value
of $\sigma_{eff}$ is given by $\sigma_{\text{eff}}=2\sigma'-\frac{2\sigma'-\sigma_{0}}{(1-\rho_{p})^{2}}$.
$\beta$, $\gamma$, $F_{max}$, $r_{h}$, $r_{c}$, $\sigma'$, and
$\sigma_{0}$ are model constants, and the values can be found in
\cite{Krassowska2007}. This equation can be written in a generalized
form as:
\begin{equation}
\dot{r}=U(r,V_{m},\rho_{p}).\label{eq: req}
\end{equation}
The equilibrium value for the pore size, $r_{eq}$, can be found by
setting the right hand side of Eq. (\ref{eq: req}) to zero: 
\begin{equation}
U(r_{eq},V_{m},\rho_{p})=0.\label{eq:req_2}
\end{equation}
In general, the dependence of $r_{eq}$ on $\rho_{p}$ is weak. On
the other hand, its dependence on $V_{m}$ exhibits an interesting
pitchfork bifurcation, which is shown in Fig. \ref{fig:bifurcation}a,
for an exemplary value of $\rho_{p}=2\times10^{-3}$. 

The coupled dynamics between $r_{eq}$ and $V_{m}$ in the initial
charging and permeabilization processes are illustrated in Figs. \ref{fig:bifurcation}a
and b, where the four stages are denoted by I-IV. In Stage I, the
initial charging stage, the membrane is near-impermeable with sub-nanometer
pores, and $V_{m}$ grows rapidly via capacitive charging (see Eq.
(\ref{eq:Ohmic})). Once $V_{m}$ reaches $V_{m}^{crit}$, the bifurcation
point, pores begin to expand significantly. Due to the consequent
``jump'' in $g_{m}$ (Stage II), $V_{m}$ eventually has to decrease
to satisfy the current continuity condition (Stage III), until the
upper branch of the bifurcation is reached. $V_{m}$ continues to
decrease until it reaches the other critical point, $V_{m}^{eq}$,
where the pore size begins to recover to the lower branch (Stage IV).
The dependence of pore radius on $V_{m}$ exhibits hysteresis typical
of this type of bifurcation. The point around $V_{m}^{eq}$ is the
most interesting, because two equilibrium pore sizes exist. For this
reason, $g_{m}$ can assume a range of values for a single value of
$V_{m}$. The upper bound can be calculated by assuming all pores
are on the upper branch, $r_{eq}=10.4$ nm; the lower bound, the lower
branch, $r_{eq}=1.01$ nm. If the required steady-state value of $g_{m}$
to satisfy the current density continuity condition (Eq. (\ref{eq:Ohmic}))
lies within these two bounds, then $V_{m}$ maintains at the level
of $V_{m}^{eq}$. This phenomenon is analogous to phase change in
physics, where $V_{m}^{eq}$ assumes the role of, e.g., temperature.
For the same argument, $V_{m}^{crit}$ is another candidate for the
equilibrium value of $V_{m}$, which in general occurs when less number
of pores are locally available, e.g., at the edges of the permeabilized
regions. (See the slightly increased $V_{m}$ values near $\theta=\pi/4,\;3\pi/4$
in Fig. \ref{fig:VmGmRhop}b.) Although the above illustration is
only schematic, the full-model simulation follows this general pattern. 

We remark that this finding bears significance in that it connects
the mesoscopic ASE model with macroscopic observables. The membrane
energy model in the ASE, based on which Eq. (\ref{eq:app1}) is derived,
has a few free constants. In general, these constants cannot be directly
measured. The above analysis reveals that if pores on the membrane
do follow the bifurcation behavior with respect to the TMP, then the
critical values, namely, $V_{m}^{eq}$ and $V_{m,}^{crit}$ can be
directly observed via a fluorescence measurement similar to that by
Kinosita et al. \cite{Kinosita_1988} or Flickinger et al. \cite{Flickinger_2010}.
These values can in turn help determine the pertinent constants (such
as $\beta$, $\gamma$, and $F_{max}$) used in the model.



\bibliographystyle{model1-num-names}
\bibliography{reference}

\begin{thebibliography}{56}
\expandafter\ifx\csname natexlab\endcsname\relax\def\natexlab#1{#1}\fi
\providecommand{\bibinfo}[2]{#2}
\ifx\xfnm\relax \def\xfnm[#1]{\unskip,\space#1}\fi
\bibitem[{Sadik et~al.(2013)Sadik, Li, Shan, Shreiber, and Lin}]{Sadik13}
\bibinfo{author}{M.~M. Sadik}, \bibinfo{author}{J.~Li}, \bibinfo{author}{J.~W.
  Shan}, \bibinfo{author}{D.~I. Shreiber}, \bibinfo{author}{H.~Lin},
\newblock \bibinfo{title}{Quantification of propidium iodide delivery using
  milisecond electric pulses: Experiments},
\newblock \bibinfo{journal}{BBA Biomembranes} \bibinfo{volume}{1828}
  (\bibinfo{year}{2013}) \bibinfo{pages}{1322--1328}.
\bibitem[{Neumann et~al.(1982)Neumann, Schaefer-Ridder, Wang, and
  Hofschneider}]{Neumann_1982}
\bibinfo{author}{E.~Neumann}, \bibinfo{author}{M.~Schaefer-Ridder},
  \bibinfo{author}{Y.~Wang}, \bibinfo{author}{P.~H. Hofschneider},
\newblock \bibinfo{title}{Gene transfer into mouse lyoma cells by
  electroporation in high electric fields},
\newblock \bibinfo{journal}{The EMBO Joumal} \bibinfo{volume}{1}
  (\bibinfo{year}{1982}) \bibinfo{pages}{841--845}.
\bibitem[{Mir(2000)}]{Mir_2000}
\bibinfo{author}{L.~M. Mir},
\newblock \bibinfo{title}{Therapeutic perspectives of in vivo cell
  electropermeabilization},
\newblock \bibinfo{journal}{Bioelectrochemistry} \bibinfo{volume}{53}
  (\bibinfo{year}{2000}) \bibinfo{pages}{1--10}.
\bibitem[{Gehl(2003)}]{Gehl_2003}
\bibinfo{author}{J.~Gehl},
\newblock \bibinfo{title}{Electroporation: theory and methods, perspectives for
  drug delivery, gene therapy and research},
\newblock \bibinfo{journal}{Acta Physiol. Scand.} \bibinfo{volume}{177}
  (\bibinfo{year}{2003}) \bibinfo{pages}{437--447}.
\bibitem[{Andr\'{e} and Mir(2004)}]{Andre_2004}
\bibinfo{author}{F.~Andr\'{e}}, \bibinfo{author}{L.~M. Mir},
\newblock \bibinfo{title}{{DNA} electrotransfer: its principles and an updated
  review of its therapeutic applications},
\newblock \bibinfo{journal}{Gene Ther.} \bibinfo{volume}{11}
  (\bibinfo{year}{2004}) \bibinfo{pages}{S33--S42}.
\bibitem[{Schoenbach et~al.(2004)Schoenbach, Joshi, Kolb, Chen, Stacey,
  Blackmore, Buescher, and Beebe}]{Schoenbach_2004}
\bibinfo{author}{K.~H. Schoenbach}, \bibinfo{author}{R.~P. Joshi},
  \bibinfo{author}{J.~F. Kolb}, \bibinfo{author}{N.~Chen},
  \bibinfo{author}{M.~Stacey}, \bibinfo{author}{P.~F. Blackmore},
  \bibinfo{author}{E.~S. Buescher}, \bibinfo{author}{S.~J. Beebe},
\newblock \bibinfo{title}{Ultrashort electrical pulses open a new gateway into
  biological cells},
\newblock \bibinfo{journal}{Proc. IEEE} \bibinfo{volume}{92}
  (\bibinfo{year}{2004}) \bibinfo{pages}{1122--1136}.
\bibitem[{{Van Driessche} et~al.(2005){Van Driessche}, Ponsaerts, {Van
  Bockstaele}, {Van Tendeloo}, and Bernema}]{Driessche_2005}
\bibinfo{author}{A.~{Van Driessche}}, \bibinfo{author}{P.~Ponsaerts},
  \bibinfo{author}{D.~R. {Van Bockstaele}}, \bibinfo{author}{V.~F.~I. {Van
  Tendeloo}}, \bibinfo{author}{Z.~N. Bernema},
\newblock \bibinfo{title}{Messenger {RNA} electroporation: an efficient tool in
  immunotherapy and stem cell research},
\newblock \bibinfo{journal}{Folia Histochemica et Cytobiologica}
  \bibinfo{volume}{43} (\bibinfo{year}{2005}) \bibinfo{pages}{213--216}.
\bibitem[{Costa et~al.(2007)Costa, Dottori, Sourris, Jamshidi, Hatzistavrou,
  Davis, Azzola, Jackson, Lim, Pera, Elefanty, and Stanley}]{Costa_2007}
\bibinfo{author}{M.~Costa}, \bibinfo{author}{M.~Dottori},
  \bibinfo{author}{K.~Sourris}, \bibinfo{author}{P.~Jamshidi},
  \bibinfo{author}{T.~Hatzistavrou}, \bibinfo{author}{R.~Davis},
  \bibinfo{author}{L.~Azzola}, \bibinfo{author}{S.~Jackson},
  \bibinfo{author}{S.~M. Lim}, \bibinfo{author}{M.~Pera},
  \bibinfo{author}{A.~G. Elefanty}, \bibinfo{author}{E.~G. Stanley},
\newblock \bibinfo{title}{A method for genetic modification of human embryonic
  stem cells using electroporation},
\newblock \bibinfo{journal}{Nat. Protoc.} \bibinfo{volume}{2}
  (\bibinfo{year}{2007}) \bibinfo{pages}{792--796}.
\bibitem[{Escoffre et~al.(2009)Escoffre, Portet, Wasungu, Teissi\'{e}, Dean,
  and Rols}]{Teissie_2009}
\bibinfo{author}{J.-M. Escoffre}, \bibinfo{author}{T.~Portet},
  \bibinfo{author}{L.~Wasungu}, \bibinfo{author}{J.~Teissi\'{e}},
  \bibinfo{author}{D.~Dean}, \bibinfo{author}{M.-P. Rols},
\newblock \bibinfo{title}{What is (still not) known of the mechanism by which
  electroporation mediates gene transfer and expression in cells and tissues},
\newblock \bibinfo{journal}{Mol Biotechnol} \bibinfo{volume}{41}
  (\bibinfo{year}{2009}) \bibinfo{pages}{286--295}.
\bibitem[{Teissi\'{e} et~al.(2012)Teissi\'{e}, Escoffre, Paganin, Chabot,
  Bellard, Wasungu, Rols, and Golzio}]{Teissie_2012}
\bibinfo{author}{J.~Teissi\'{e}}, \bibinfo{author}{J.-M. Escoffre},
  \bibinfo{author}{A.~Paganin}, \bibinfo{author}{S.~Chabot},
  \bibinfo{author}{E.~Bellard}, \bibinfo{author}{L.~Wasungu},
  \bibinfo{author}{M.-P. Rols}, \bibinfo{author}{M.~Golzio},
\newblock \bibinfo{title}{Drug delivery by electropulsation: Recent
  developments in oncology},
\newblock \bibinfo{journal}{Int. J. Pharm.} \bibinfo{volume}{423}
  (\bibinfo{year}{2012}) \bibinfo{pages}{3--6}.
\bibitem[{Chang and Reese(1990)}]{Chang_1990}
\bibinfo{author}{D.~C. Chang}, \bibinfo{author}{T.~S. Reese},
\newblock \bibinfo{title}{Changes in membrane structure induced by
  electroporation as revealed by rapid-freezing electron microscopy},
\newblock \bibinfo{journal}{Biophys. J.} \bibinfo{volume}{58}
  (\bibinfo{year}{1990}) \bibinfo{pages}{1--12}.
\bibitem[{Wilhelm et~al.(1993)Wilhelm, Winterhalter, Zimmermann, and
  Benz}]{Wilhelm_1993}
\bibinfo{author}{C.~Wilhelm}, \bibinfo{author}{M.~Winterhalter},
  \bibinfo{author}{U.~Zimmermann}, \bibinfo{author}{R.~Benz},
\newblock \bibinfo{title}{Kinetics of pore size during irreversible electrical
  breakdown of lipid bilayer membranes},
\newblock \bibinfo{journal}{Biophys. J.} \bibinfo{volume}{64}
  (\bibinfo{year}{1993}) \bibinfo{pages}{121--128}.
\bibitem[{Spassova et~al.(1994)Spassova, Tsoneva, Petrov, Petkova, and
  Neumann}]{Spassova_1994}
\bibinfo{author}{M.~Spassova}, \bibinfo{author}{I.~Tsoneva},
  \bibinfo{author}{A.~G. Petrov}, \bibinfo{author}{J.~I. Petkova},
  \bibinfo{author}{E.~Neumann},
\newblock \bibinfo{title}{Dip patch clamp currents suggest electrodiffusive
  transport of the polyelectrolyte dna through lipid bilayers},
\newblock \bibinfo{journal}{Biophys. Chem.} \bibinfo{volume}{52}
  (\bibinfo{year}{1994}) \bibinfo{pages}{267--274}.
\bibitem[{Djuzenova et~al.(1996)Djuzenova, Zimmermann, Frank, Sukhorukov,
  Richter, and Fuhr}]{Djuzenova_1996}
\bibinfo{author}{C.~S. Djuzenova}, \bibinfo{author}{U.~Zimmermann},
  \bibinfo{author}{H.~Frank}, \bibinfo{author}{V.~L. Sukhorukov},
  \bibinfo{author}{E.~Richter}, \bibinfo{author}{G.~Fuhr},
\newblock \bibinfo{title}{Effect of medium conductivity and composition on the
  uptake of propidium iodide into electropermeabilized myeloma cells},
\newblock \bibinfo{journal}{Biochim. Biophys. Acta} \bibinfo{volume}{1284}
  (\bibinfo{year}{1996}) \bibinfo{pages}{143--152}.
\bibitem[{Wegner et~al.(2011)Wegner, Flickinger, Eing, Bergh\"{o}fer,
  Hohenberger, Frey, and Nick}]{Wegner_2011}
\bibinfo{author}{L.~H. Wegner}, \bibinfo{author}{B.~Flickinger},
  \bibinfo{author}{C.~Eing}, \bibinfo{author}{T.~Bergh\"{o}fer},
  \bibinfo{author}{P.~Hohenberger}, \bibinfo{author}{W.~Frey},
  \bibinfo{author}{P.~Nick},
\newblock \bibinfo{title}{A patch clamp study on the electro-permeabilization
  of higher plant cells: Supra-physiological voltages induce a
  high-conductance, {K}+ selective state of the plasma membrane},
\newblock \bibinfo{journal}{Biochim. Biophys. Acta} \bibinfo{volume}{1808}
  (\bibinfo{year}{2011}) \bibinfo{pages}{1728--1736}.
\bibitem[{Napotnik et~al.(2012)Napotnik, Wu, Gundersen, Miklav\v{c}i\v{c}, and
  Vernier}]{Napotnik_2012}
\bibinfo{author}{T.~B. Napotnik}, \bibinfo{author}{Y.-H. Wu},
  \bibinfo{author}{M.~A. Gundersen}, \bibinfo{author}{D.~Miklav\v{c}i\v{c}},
  \bibinfo{author}{P.~T. Vernier},
\newblock \bibinfo{title}{Nanosecond electric pulses cause mitochondrial
  membrane permeabilization in jurkat cells},
\newblock \bibinfo{journal}{Bioelectromagnetics} \bibinfo{volume}{33}
  (\bibinfo{year}{2012}) \bibinfo{pages}{257--264}.
\bibitem[{Kramar et~al.(2012)Kramar, Delemotte, Lebar, Kotulska, Tarek, and
  Miklav\v{c}i\v{c}}]{Krama_2012}
\bibinfo{author}{P.~Kramar}, \bibinfo{author}{L.~Delemotte},
  \bibinfo{author}{A.~M. Lebar}, \bibinfo{author}{M.~Kotulska},
  \bibinfo{author}{M.~Tarek}, \bibinfo{author}{D.~Miklav\v{c}i\v{c}},
\newblock \bibinfo{title}{Molecular-level characterization of lipid membrane
  electroporation using linearly rising current},
\newblock \bibinfo{journal}{Journal of Membrane Biology} \bibinfo{volume}{245}
  (\bibinfo{year}{2012}) \bibinfo{pages}{651--659}.
\bibitem[{Barnett and Weaver(1991)}]{Barnett_1991}
\bibinfo{author}{A.~Barnett}, \bibinfo{author}{J.~C. Weaver},
\newblock \bibinfo{title}{Electroporation: a unified, quantitative theory of
  reversible electrical breakdown and mechanical rupture in artificial planar
  bilayer membranes},
\newblock \bibinfo{journal}{Bioelectrochem. Bioenerg.} \bibinfo{volume}{25}
  (\bibinfo{year}{1991}) \bibinfo{pages}{163--182}.
\bibitem[{Weaver and Chizmadzhev(1996)}]{Weaver_1996}
\bibinfo{author}{J.~C. Weaver}, \bibinfo{author}{Y.~A. Chizmadzhev},
\newblock \bibinfo{title}{Theory of electroporation: A review},
\newblock \bibinfo{journal}{Bioelectrochem. Bioenerg.} \bibinfo{volume}{41}
  (\bibinfo{year}{1996}) \bibinfo{pages}{135--160}.
\bibitem[{Neu and Krassowska(1999)}]{Neu_1999}
\bibinfo{author}{J.~C. Neu}, \bibinfo{author}{W.~Krassowska},
\newblock \bibinfo{title}{Asymptotic model of electroporation},
\newblock \bibinfo{journal}{Phys. Rev. E} \bibinfo{volume}{59}
  (\bibinfo{year}{1999}) \bibinfo{pages}{3471--3482}.
\bibitem[{Leontiadou et~al.(2004)Leontiadou, Mark, and
  Marrink}]{Leontiadou_2004}
\bibinfo{author}{H.~Leontiadou}, \bibinfo{author}{A.~E. Mark},
  \bibinfo{author}{S.~J. Marrink},
\newblock \bibinfo{title}{Molecular dynamics simulations of hydrophilic pores
  in lipid bilayers},
\newblock \bibinfo{journal}{Biophys. J.} \bibinfo{volume}{86}
  (\bibinfo{year}{2004}) \bibinfo{pages}{2156--2164}.
\bibitem[{Tarek(2005)}]{Tarek_2005}
\bibinfo{author}{M.~Tarek},
\newblock \bibinfo{title}{Membrane electroporation: A molecular dynamics
  simulation},
\newblock \bibinfo{journal}{Biophys. J.} \bibinfo{volume}{88}
  (\bibinfo{year}{2005}) \bibinfo{pages}{4045--4053}.
\bibitem[{Levine and Vernier(2010)}]{Levine_2010}
\bibinfo{author}{Z.~A. Levine}, \bibinfo{author}{P.~T. Vernier},
\newblock \bibinfo{title}{Life cycle of an electropore: Field-dependent and
  field-independent steps in pore creation and annihilation},
\newblock \bibinfo{journal}{J Membrane Biol} \bibinfo{volume}{236}
  (\bibinfo{year}{2010}) \bibinfo{pages}{27--36}.
\bibitem[{Li and Lin(2010)}]{Jianbo_2010}
\bibinfo{author}{J.~Li}, \bibinfo{author}{H.~Lin},
\newblock \bibinfo{title}{The current-voltage relation for electropores with
  conductivity gradients},
\newblock \bibinfo{journal}{Biomicrofluidics} \bibinfo{volume}{4}
  (\bibinfo{year}{2010}) \bibinfo{pages}{013206}.
\bibitem[{Sukharev et~al.(1992)Sukharev, Klenchin, Serov, Chernomordik, and
  Chizmadzhev}]{sukharev_1992}
\bibinfo{author}{S.~I. Sukharev}, \bibinfo{author}{V.~A. Klenchin},
  \bibinfo{author}{S.~M. Serov}, \bibinfo{author}{L.~V. Chernomordik},
  \bibinfo{author}{Y.~A. Chizmadzhev},
\newblock \bibinfo{title}{Electroporation and electrophoretic {DNA} transfer
  into cells: the effect of {DNA} interaction with electropores},
\newblock \bibinfo{journal}{Biophys. J.} \bibinfo{volume}{63}
  (\bibinfo{year}{1992}) \bibinfo{pages}{1320--1327}.
\bibitem[{Prausnitz et~al.(1995)Prausnitz, Corbett, Gimm, Golan, Langer, and
  Weaver}]{prausnitz_1995}
\bibinfo{author}{M.~R. Prausnitz}, \bibinfo{author}{J.~D. Corbett},
  \bibinfo{author}{J.~A. Gimm}, \bibinfo{author}{D.~E. Golan},
  \bibinfo{author}{R.~Langer}, \bibinfo{author}{J.~C. Weaver},
\newblock \bibinfo{title}{Millisecond measurement of transport during and after
  an electroporation pulse},
\newblock \bibinfo{journal}{Biophys. J.} \bibinfo{volume}{68}
  (\bibinfo{year}{1995}) \bibinfo{pages}{1864--1870}.
\bibitem[{Mir et~al.(1999)Mir, Bureau, Gehl, Rangara, Rouyi, Caillaud, Delaere,
  Branelleci, Schwartz, and Scherman}]{Mir_1998}
\bibinfo{author}{L.~M. Mir}, \bibinfo{author}{M.~F. Bureau},
  \bibinfo{author}{J.~Gehl}, \bibinfo{author}{R.~Rangara},
  \bibinfo{author}{D.~Rouyi}, \bibinfo{author}{J.-M. Caillaud},
  \bibinfo{author}{P.~Delaere}, \bibinfo{author}{D.~Branelleci},
  \bibinfo{author}{B.~Schwartz}, \bibinfo{author}{D.~Scherman},
\newblock \bibinfo{title}{High-efficiency gene transfer into skeletal muscle
  mediated by electric pulses},
\newblock \bibinfo{journal}{Proc. Natl. Acad. Sci.} \bibinfo{volume}{96}
  (\bibinfo{year}{1999}) \bibinfo{pages}{4262--4267}.
\bibitem[{Rols and Teissi\'{e}(1998)}]{Rols_1998}
\bibinfo{author}{M.-P. Rols}, \bibinfo{author}{J.~Teissi\'{e}},
\newblock \bibinfo{title}{Electropermeabilization of mammalian cells to
  macromolecules: Control by pulse duration},
\newblock \bibinfo{journal}{Biophys. J.} \bibinfo{volume}{75}
  (\bibinfo{year}{1998}) \bibinfo{pages}{1415--1423}.
\bibitem[{Faurie et~al.(2004)Faurie, Phez, Golzio, Vossen, Lesbordes, Delteil,
  Teissi\'{e}, and Rols}]{Faurie_2004}
\bibinfo{author}{C.~Faurie}, \bibinfo{author}{E.~Phez},
  \bibinfo{author}{M.~Golzio}, \bibinfo{author}{C.~Vossen},
  \bibinfo{author}{J.-C. Lesbordes}, \bibinfo{author}{C.~Delteil},
  \bibinfo{author}{J.~Teissi\'{e}}, \bibinfo{author}{M.-P. Rols},
\newblock \bibinfo{title}{Effect of electric field vectoriality on electrically
  mediated gene delivery in mammalian cells},
\newblock \bibinfo{journal}{Biochim. Biophys. Acta} \bibinfo{volume}{1665}
  (\bibinfo{year}{2004}) \bibinfo{pages}{92--100}.
\bibitem[{\v{S}atkauskas et~al.(2002)\v{S}atkauskas, Bureau, Puc, Mahfoudi,
  Scherman, Miklav\v{c}i\v{c}, and Mir}]{Satkauskas_2002}
\bibinfo{author}{S.~\v{S}atkauskas}, \bibinfo{author}{M.~F. Bureau},
  \bibinfo{author}{M.~Puc}, \bibinfo{author}{A.~Mahfoudi},
  \bibinfo{author}{D.~Scherman}, \bibinfo{author}{D.~Miklav\v{c}i\v{c}},
  \bibinfo{author}{L.~M. Mir},
\newblock \bibinfo{title}{Mechanisms of in vivo {DNA} electrotransfer:
  Respective contributions of cell electropermeabilization and {DNA}
  electrophoresis},
\newblock \bibinfo{journal}{Molecular Therapy} \bibinfo{volume}{5}
  (\bibinfo{year}{2002}) \bibinfo{pages}{133--140}.
\bibitem[{\v{S}atkauskas et~al.(2005)\v{S}atkauskas, Andr\'{e}, Bureau,
  Scherman, Miklav\v{c}i\v{c}, and Mir}]{Satkauskas_2005}
\bibinfo{author}{S.~\v{S}atkauskas}, \bibinfo{author}{F.~Andr\'{e}},
  \bibinfo{author}{M.~F. Bureau}, \bibinfo{author}{D.~Scherman},
  \bibinfo{author}{D.~Miklav\v{c}i\v{c}}, \bibinfo{author}{L.~M. Mir},
\newblock \bibinfo{title}{Electrophoretic component of electric pulses
  determines the efficacy of in vivo {DNA} electrotransfer},
\newblock \bibinfo{journal}{Hum. Gene Ther.} \bibinfo{volume}{16}
  (\bibinfo{year}{2005}) \bibinfo{pages}{1194--1201}.
\bibitem[{Pav\v{s}elj and Pr\'{e}at(2005)}]{Pavselj_2005}
\bibinfo{author}{N.~Pav\v{s}elj}, \bibinfo{author}{V.~Pr\'{e}at},
\newblock \bibinfo{title}{{DNA} electrotransfer into the skin using a
  combination of one high- and one low-voltage pulse},
\newblock \bibinfo{journal}{J. Controlled Release} \bibinfo{volume}{106}
  (\bibinfo{year}{2005}) \bibinfo{pages}{407--415}.
\bibitem[{Liu et~al.(2006)Liu, Heston, Shollenberger, Sun, Mickle, Lovell, and
  Huang}]{Liu_2006}
\bibinfo{author}{F.~Liu}, \bibinfo{author}{S.~Heston}, \bibinfo{author}{L.~M.
  Shollenberger}, \bibinfo{author}{B.~Sun}, \bibinfo{author}{M.~Mickle},
  \bibinfo{author}{M.~Lovell}, \bibinfo{author}{L.~Huang},
\newblock \bibinfo{title}{Mechanism of in vivo {DNA} transport into cells by
  electroporation: electrophoresis across the plasma membrane may not be
  involved},
\newblock \bibinfo{journal}{Journal of Gene Medicine} \bibinfo{volume}{8}
  (\bibinfo{year}{2006}) \bibinfo{pages}{353--361}.
\bibitem[{Pucihar et~al.(2008)Pucihar, Kotnik, Miklav\v{c}i\v{c}, and
  Teissi\'{e}}]{Pucihar_2008}
\bibinfo{author}{G.~Pucihar}, \bibinfo{author}{T.~Kotnik},
  \bibinfo{author}{D.~Miklav\v{c}i\v{c}}, \bibinfo{author}{J.~Teissi\'{e}},
\newblock \bibinfo{title}{Kinetics of transmembrane transport of small
  molecules into electropermeabilized cells},
\newblock \bibinfo{journal}{Biophys. J.} \bibinfo{volume}{95}
  (\bibinfo{year}{2008}) \bibinfo{pages}{2837--2848}.
\bibitem[{Kandu\v{s}er et~al.(2009)Kandu\v{s}er, Miklav\v{c}i\v{c}, and
  Pavlin}]{Kanduser_2009}
\bibinfo{author}{M.~Kandu\v{s}er}, \bibinfo{author}{D.~Miklav\v{c}i\v{c}},
  \bibinfo{author}{M.~Pavlin},
\newblock \bibinfo{title}{Mechanisms involved in gene electrotransfer using
  high- and low-voltage pulses - an in vitro study},
\newblock \bibinfo{journal}{Bioelectrochemistry} \bibinfo{volume}{74}
  (\bibinfo{year}{2009}) \bibinfo{pages}{265--271}.
\bibitem[{Miklav\v{c}i\v{c} and Towhidi(2010)}]{Miklavcic_2010}
\bibinfo{author}{D.~Miklav\v{c}i\v{c}}, \bibinfo{author}{L.~Towhidi},
\newblock \bibinfo{title}{Numerical study of the electroporation pulse shape
  effect on molecular uptake of biological cells},
\newblock \bibinfo{journal}{Radiol Oncol} \bibinfo{volume}{44}
  (\bibinfo{year}{2010}) \bibinfo{pages}{34--41}.
\bibitem[{Pavlin et~al.(2010)Pavlin, Flisar, and Kandu\v{s}er}]{Pavlin_2010_2}
\bibinfo{author}{M.~Pavlin}, \bibinfo{author}{K.~Flisar},
  \bibinfo{author}{M.~Kandu\v{s}er},
\newblock \bibinfo{title}{The role of electrophoresis in gene electrotransfer},
\newblock \bibinfo{journal}{J. Membrane Biol.} \bibinfo{volume}{236}
  (\bibinfo{year}{2010}) \bibinfo{pages}{75--79}.
\bibitem[{Smith and Weaver(2011)}]{Smith_2012}
\bibinfo{author}{K.~C. Smith}, \bibinfo{author}{J.~C. Weaver},
\newblock \bibinfo{title}{Transmembrane molecular transport during versus after
  extremely large, nanosecond electric pulses},
\newblock \bibinfo{journal}{Biochem. Biophys. Res. Commun.}
  \bibinfo{volume}{412} (\bibinfo{year}{2011}) \bibinfo{pages}{8--12}.
\bibitem[{V\'{a}squez et~al.(2012)V\'{a}squez, Gehl, and
  Hermann}]{Vasquez_2012}
\bibinfo{author}{J.~L. V\'{a}squez}, \bibinfo{author}{J.~Gehl},
  \bibinfo{author}{G.~G. Hermann},
\newblock \bibinfo{title}{Electroporation enhances mitomycin {C} cytotoxicity
  on {T}24 bladder cancer cell line: A potential improvement of intravesical
  chemotherapy in bladder cancer},
\newblock \bibinfo{journal}{Bioelectrochemistry} \bibinfo{volume}{88}
  (\bibinfo{year}{2012}) \bibinfo{pages}{127--133}.
\bibitem[{Zimmermann et~al.(1990)Zimmermann, Schnettler, Kl\"{o}ck, and
  Watzka}]{Zimmermann_1990}
\bibinfo{author}{U.~Zimmermann}, \bibinfo{author}{R.~Schnettler},
  \bibinfo{author}{G.~Kl\"{o}ck}, \bibinfo{author}{H.~Watzka},
\newblock \bibinfo{title}{Mechanisms of electrostimulated uptake of
  macromolecules into living cells},
\newblock \bibinfo{journal}{Naturwissenschaften} \bibinfo{volume}{77}
  (\bibinfo{year}{1990}) \bibinfo{pages}{543--545}.
\bibitem[{Wu and Yuan(2011)}]{Wu_2011}
\bibinfo{author}{M.~Wu}, \bibinfo{author}{F.~Yuan},
\newblock \bibinfo{title}{Membrane binding of plasmid {DNA} and endocytic
  pathways are involved in electrotransfection of mammalian cells},
\newblock \bibinfo{journal}{PLoS ONE} \bibinfo{volume}{6}
  (\bibinfo{year}{2011}) \bibinfo{pages}{e20923}.
\bibitem[{Lin et~al.(2011)Lin, Chang, and Lee}]{Lin_2011}
\bibinfo{author}{R.~Lin}, \bibinfo{author}{D.~C. Chang}, \bibinfo{author}{Y.-K.
  Lee},
\newblock \bibinfo{title}{Single-cell electroendocytosis on a micro chip using
  in situ fluorescence microscopy},
\newblock \bibinfo{journal}{Biomed. Microdevices} \bibinfo{volume}{13}
  (\bibinfo{year}{2011}) \bibinfo{pages}{1063--1073}.
\bibitem[{Rosazza et~al.(2012)Rosazza, Phez, Escoffre, C\'{e}zanne, Zumbusch,
  and Rols}]{Rosazza_2012}
\bibinfo{author}{C.~Rosazza}, \bibinfo{author}{E.~Phez}, \bibinfo{author}{J.-M.
  Escoffre}, \bibinfo{author}{L.~C\'{e}zanne}, \bibinfo{author}{A.~Zumbusch},
  \bibinfo{author}{M.-P. Rols},
\newblock \bibinfo{title}{Cholesterol implications in plasmid {DNA}
  electrotransfer: Evidence for the involvement of endocytotic pathways},
\newblock \bibinfo{journal}{Int. J. Pharm.} \bibinfo{volume}{423}
  (\bibinfo{year}{2012}) \bibinfo{pages}{134--143}.
\bibitem[{M\"{u}ller et~al.(2001)M\"{u}ller, Sukhorukov, and
  Zimmermann}]{Muller_2001}
\bibinfo{author}{K.~J. M\"{u}ller}, \bibinfo{author}{V.~L. Sukhorukov},
  \bibinfo{author}{U.~Zimmermann},
\newblock \bibinfo{title}{Reversible electropermeabilization of mammalian cells
  by high-intensity, ultra-short pulses of submicrosecond duration},
\newblock \bibinfo{journal}{J. Membrane Biol.} \bibinfo{volume}{184}
  (\bibinfo{year}{2001}) \bibinfo{pages}{161--170}.
\bibitem[{Neu and Krassowska(2003)}]{Neu_2003}
\bibinfo{author}{J.~C. Neu}, \bibinfo{author}{W.~Krassowska},
\newblock \bibinfo{title}{Modeling postshock evolution of large electropores},
\newblock \bibinfo{journal}{Phys. Rev. E} \bibinfo{volume}{67}
  (\bibinfo{year}{2003}) \bibinfo{pages}{021915}.
\bibitem[{Krassowska and Filev(2007)}]{Krassowska2007}
\bibinfo{author}{W.~Krassowska}, \bibinfo{author}{P.~D. Filev},
\newblock \bibinfo{title}{Modeling electroporation in a single cell},
\newblock \bibinfo{journal}{Biophys. J.} \bibinfo{volume}{92}
  (\bibinfo{year}{2007}) \bibinfo{pages}{404--417}.
\bibitem[{Li and Lin(2011)}]{Jianbo_2011}
\bibinfo{author}{J.~Li}, \bibinfo{author}{H.~Lin},
\newblock \bibinfo{title}{Numerical simulation of molecular uptake via
  electroporation},
\newblock \bibinfo{journal}{Bioelectrochemistry} \bibinfo{volume}{82}
  (\bibinfo{year}{2011}) \bibinfo{pages}{10--21}.
\bibitem[{Li et~al.(2012)Li, Tan, Yu, and Lin}]{Jianbo_2012}
\bibinfo{author}{J.~Li}, \bibinfo{author}{W.~Tan}, \bibinfo{author}{M.~Yu},
  \bibinfo{author}{H.~Lin},
\newblock \bibinfo{title}{The effect of extracellular conductivity on
  electroporation mediated molecular delivery},
\newblock \bibinfo{journal}{BBA Biomembranes} \bibinfo{volume}{1828}
  (\bibinfo{year}{2012}) \bibinfo{pages}{461--470}.
\bibitem[{Sadik et~al.(2013)Sadik, Yu, Shan, Shreiber, and Lin}]{Sadik_2013_2}
\bibinfo{author}{M.~M. Sadik}, \bibinfo{author}{M.~Yu}, \bibinfo{author}{J.~W.
  Shan}, \bibinfo{author}{D.~I. Shreiber}, \bibinfo{author}{H.~Lin},
\newblock \bibinfo{title}{Scaling relationship and optimization of double-pulse
  electroporation},
\newblock \bibinfo{journal}{Biophys. J., accepted}  (\bibinfo{year}{2013}).
\bibitem[{Vlahovska et~al.(2009)Vlahovska, Graci\`{a}, Aranda-Espinoza, and
  Dimova}]{Vlahovska_2009}
\bibinfo{author}{P.~M. Vlahovska}, \bibinfo{author}{R.~S. Graci\`{a}},
  \bibinfo{author}{S.~Aranda-Espinoza}, \bibinfo{author}{R.~Dimova},
\newblock \bibinfo{title}{Electrohydrodynamic model of vesicle deformation in
  alternating electric fields},
\newblock \bibinfo{journal}{Biophys. J.} \bibinfo{volume}{96}
  (\bibinfo{year}{2009}) \bibinfo{pages}{4789--4803}.
\bibitem[{Sadik et~al.(2011)Sadik, Li, Shan, Shreiber, and Lin}]{Sadik_2011}
\bibinfo{author}{M.~M. Sadik}, \bibinfo{author}{J.~Li}, \bibinfo{author}{J.~W.
  Shan}, \bibinfo{author}{D.~I. Shreiber}, \bibinfo{author}{H.~Lin},
\newblock \bibinfo{title}{Vesicle deformation and poration under strong dc
  electric fields},
\newblock \bibinfo{journal}{Phy. Rev. E.} \bibinfo{volume}{83}
  (\bibinfo{year}{2011}) \bibinfo{pages}{066316}.
\bibitem[{Suzuki et~al.(2011)Suzuki, Ramos, Ribeiro, Cazarolli, Silva, Leite,
  and Marques}]{Suzuki_2011}
\bibinfo{author}{D.~O.~H. Suzuki}, \bibinfo{author}{A.~Ramos},
  \bibinfo{author}{M.~C.~M. Ribeiro}, \bibinfo{author}{L.~H. Cazarolli},
  \bibinfo{author}{F.~R. M.~B. Silva}, \bibinfo{author}{L.~D. Leite},
  \bibinfo{author}{J.~L.~B. Marques},
\newblock \bibinfo{title}{Theoretical and experimental analysis of
  electroporated membrane conductance in cell suspension},
\newblock \bibinfo{journal}{IEEE Trans. Biomed. Eng.} \bibinfo{volume}{58}
  (\bibinfo{year}{2011}) \bibinfo{pages}{3310--3318}.
\bibitem[{He et~al.(2008)He, Chang, and Lee}]{He_2008}
\bibinfo{author}{H.~He}, \bibinfo{author}{D.~C. Chang}, \bibinfo{author}{Y.-K.
  Lee},
\newblock \bibinfo{title}{Nonlinear current response of micro electroporation
  and resealing dynamics for human cancer cells},
\newblock \bibinfo{journal}{Bioelectrochemistry} \bibinfo{volume}{72}
  (\bibinfo{year}{2008}) \bibinfo{pages}{161--168}.
\bibitem[{{Kinosita Jr.} et~al.(1988){Kinosita Jr.}, Ashikawa, Saita,
  Yoshimura, Itoh, Nagayama, and Ikegami}]{Kinosita_1988}
\bibinfo{author}{K.~{Kinosita Jr.}}, \bibinfo{author}{I.~Ashikawa},
  \bibinfo{author}{N.~Saita}, \bibinfo{author}{H.~Yoshimura},
  \bibinfo{author}{H.~Itoh}, \bibinfo{author}{K.~Nagayama},
  \bibinfo{author}{A.~Ikegami},
\newblock \bibinfo{title}{Electroporation of cell membrane visualized under a
  pulsed-laser fluorescence microscope},
\newblock \bibinfo{journal}{Biophys. J.} \bibinfo{volume}{53}
  (\bibinfo{year}{1988}) \bibinfo{pages}{1015--1019}.
\bibitem[{Flickinger et~al.(2010)Flickinger, Bergh\"{o}fer, Hohenberger, Eing,
  and Frey}]{Flickinger_2010}
\bibinfo{author}{B.~Flickinger}, \bibinfo{author}{T.~Bergh\"{o}fer},
  \bibinfo{author}{P.~Hohenberger}, \bibinfo{author}{C.~Eing},
  \bibinfo{author}{W.~Frey},
\newblock \bibinfo{title}{Transmembrane potential measurements on plant cells
  using the voltage-sensitive dye {ANNINE}-6},
\newblock \bibinfo{journal}{Protoplasma} \bibinfo{volume}{247}
  (\bibinfo{year}{2010}) \bibinfo{pages}{3--12}.
\bibitem[{Freeman et~al.(1994)Freeman, Wang, and Weaver}]{Freeman_1994}
\bibinfo{author}{S.~A. Freeman}, \bibinfo{author}{M.~A. Wang},
  \bibinfo{author}{J.~C. Weaver},
\newblock \bibinfo{title}{Theory of electroporation of planar bilayer
  membranes: Predictions of the aqueous area, change in capacitance, and
  pore-pore separation},
\newblock \bibinfo{journal}{Biophys. J.} \bibinfo{volume}{67}
  (\bibinfo{year}{1994}) \bibinfo{pages}{42--56}.

\end{thebibliography}







\end{document}